\documentclass[preprint]{aastex63}
\usepackage{graphicx}
\usepackage{bm}
\usepackage{verbatim}

\usepackage{placeins}

\usepackage{amsmath}

\definecolor{forestgreen}{rgb}{0.10, 0.50, 0.10}
\newcommand{\peijin}[1]{{#1}}
\received{October 19, 2020}
\revised{\today}
\accepted{ xxx}
\submitjournal{xxx}

\shorttitle{Wave propagation effects on the source of solar radio bursts}
\shortauthors{Zhang et al.}

\begin{document}

\title{Parametric simulation studies on the wave propagation of solar radio emission: \\
the source size, duration, and position}

\correspondingauthor{ChuanBing Wang, Eduard P. Kontar}
\email{cbwang@ustc.edu.cn}
\email{eduard.kontar@glasgow.ac.uk}
\author[0000-0001-6855-5799]{PeiJin Zhang}
\affil{CAS Key Laboratory of Geospace Environment,
	 School of Earth and Space Sciences, \\
	 University of Science and Technology of China (USTC),
	  Hefei, Anhui 230026, China}
\affiliation{CAS Center for the Excellence in Comparative Planetology, USTC, Hefei, Anhui 230026, China}

\author[0000-0001-6252-5580]{ChuanBing Wang}
\affiliation{CAS Key Laboratory of Geospace Environment,
	School of Earth and Space Sciences, \\
	University of Science and Technology of China (USTC),
	Hefei, Anhui 230026, China}
\affiliation{CAS Center for the Excellence in Comparative Planetology, USTC, Hefei, Anhui 230026, China}


\author[0000-0002-8078-0902]{Eduard P. Kontar}
\affiliation{School of Physics and Astronomy, University of Glasgow, Glasgow G12 8QQ, UK}



\begin{abstract}
The observed features of the radio sources indicate complex propagation effects embedded in the waves of solar radio bursts. 
In this work, we perform ray-tracing simulations on radio wave transport in the corona and interplanetary region with anisotropic electron density fluctuations. \peijin{For the first time, } 
the variation of the apparent source size, burst duration, and source position \peijin{of both fundamental emission and harmonic emission} at frequency 35\,MHz are simulated as the function of the anisotropic  parameter $\alpha$ and the angular scattering rate coefficient $\eta =\epsilon^2/h_0$, where $\epsilon^2={\langle \delta n^2\rangle}/{n^2}$ is the density fluctuation level and $h_0$ is its correlation length near the wave exciting site.   
It is found that isotropic fluctuations produce a much larger decay time than a highly anisotropic fluctuation \peijin{for fundamental emission}. By comparing the observed duration and source size with the simulation results in the parameter space, we can estimate the scattering coefficient and the anisotropy parameter  $\eta = 8.9\times 10^{-5}\, \mathrm{km^{-1}}$ and  $\alpha = 0.719$  with point pulse source assumption. Position offsets due to wave scattering and refraction can produce the co-spatial of fundamental and harmonic waves in observations \peijin{of some type III radio bursts}. 
The visual speed due to the wave propagation effect can reach 1.5\,$c$ for  $\eta = 2.4\times 10^{-4}\, \mathrm{km^{-1}}$ and $\alpha=0.2$ for fundamental emission in the sky plane, \peijin{accompanying with large expansion rate of the source size}. 
The visual speed direction is mostly identical to the offset direction, thus, for the observation aiming at obtaining the source position, the source centroid at the starting point is closer to the wave excitation point. 
\end{abstract}


\keywords{solar radio burst --- 
source size and position --- wave propagation effects}


\section{Introduction}

The imaging and spectroscopy observations of the solar radio bursts can provide information on the non-thermal electrons associated with the transient energy release in the solar active region and parameters of the background plasma.  For example, 
\cite{mccauley2018densities} inspected the background density of the solar corona with the interferometric imaging of Type III radio bursts in the frequency range of 80-240\,MHz.
High-cadence radio imaging spectroscopy shows evidence of \peijin{the existence of} particle acceleration by solar flare termination shock \citep{chen2015Science,Yu2020biflow}.
The combined observations of radio imaging and extreme ultraviolet /white-light imaging indicate that the particle acceleration occurs at the flank of coronal mass ejection shock \citep{chrysaphi2018cme,morosan2019multiple,chen2014solar}.  The imaging spectroscopy study with LOw Frequency Array (LOFAR) reveals that the velocity dispersion of the electron beams is a key factor for the duration of type III radio burst \citep{zhang2019source}.

The corona plasma is an in-homogeneous refractive media for solar radio waves, so the refraction and scattering can cause the deformation of the observed radio source including the expansion of the source size, the offset of the visual source position from the wave generation position \citep{wild1959transverse,kontar2017imaging,bisoi2018decimetric}, and the duration broadening in the dynamic spectrum \citep{zhang2019source}.
Generally, the refraction can cause the inward offset of the visual source from the position of wave excitation \citep{mann2018tracking}. The visual source could be shifted outward if the scattering is considered \citep{stewart1972relative,stewart1976source,arzner1999radiowave,kontar2019anisotropic}. 
For the wave of solar radio burst with fundamental-harmonic (F-H) pair structure, the H-emission is generated at a higher height than the F-emission with the same frequency due to the plasma emission mechanism, while 
imaging result indicates that the source of F and H emission \peijin{can} have a similar position \peijin{for some bursts} \citep{stewart1972relative,dulk1980position}. This seems to indicate that the waves of F-emission and H-emission have experienced different amounts of refraction and scattering. 


For a given frequency, the Langmuir wave excitation responsible for type III emission is limited to the cross-section of the guiding magnetic field line for the electrons and the layer with local plasma frequency close to $1$ or $1/2$ times of the radio wave frequency. The structure of the magnetic field and the distribution of the background density is stable within the time scale of electron beam transit time. Thus the excitation site at a fixed frequency is stable in both size and position. While the visual velocity and expansion rate of the observed sources can be very large.  \cite{kontar2017imaging} observed 1/4\,$c$ radial speed with LOFAR beamformed observation of the fundamental part of a type IIIb striae, where $c$ is the speed of light. \cite{kuznetsov2020radio} observed 1/3\,$c$ at 32\,MHz in a drift pair radio burst.
The interferometric imaging with the remote baseline of LOFAR finds about 4\,$c$ visual speed for the fundamental part of a type III-IIIb pair event source at 26\,MHz \citep{zhang2020interferometric}. The observed source size of the radio burst is also considerably larger than the size estimated from the bandwidth of the striae \citep{kontar2017imaging,zhang2020interferometric}.
The fast variations of the source is believed to be due to the wave propagation effects. 


The ray-tracing simulation is an effective method to investigate the wave propagation effects, which can help us interpreting the imaging and spectroscopy observation and diagnosing the solar corona properties from the observation results \citep{steinberg1971coronal,riddle1974observation}. The ray-tracing method is introduced to the solar radio study by \cite{fokker1965coronal} to investigate the source size of type I radio burst due to the scattering effect. \cite{bougeret1977new} indicated that the over-dense fiber structures in the coronal should be considered to understand the position of the radio source, the moving bursts, and the great variety of space-time shapes observed within the same storm center, and \cite{robinson1983scattering} introduced fiber in-homogeneity into the ray tracing simulation and find that the in-homogeneity of the medium can account for the source displacement.
It has been pointed out that the anisotropic scattering of the wave due to the statistical inhomogeneity is essential to interpret the observed radio source properties \citep{arzner1999radiowave,kontar2019anisotropic,bian2019fokker}. 
Recently, \cite{kontar2019anisotropic} developed a model to include the anisotropic scattering of the wave in the ray tracing process. \peijin{It was found that the  duration of type III radio burst decrease with the anisotropy level of the background, and the anisotropic density fluctuations are required to account for the source sizes and decay times simulaneously.} The theory of anisotropic scattering has been successfully used to interpret the source properties of the drift pair bursts\citep{kuznetsov2019first,kuznetsov2020radio}. 

These case studies of observation with the corresponding simulation have shown that the ray-tracing is an effective method to analyze the wave propagation effect, while the detailed dependency relation between the observed source characteristics and the density fluctuation properties of the medium is still unclear. A further understanding of the physical process behind the visual source variation requires more detailed simulation-observation comparison study.


In this work, \peijin{for the first time,} we performed a large set of Monte Carlo simulations of the radio wave propagation by ray tracing to \peijin{explore a parameter-space study on} the radio source properties with various background plasma parameters. The paper is arranged as flows: in Section 2, the model used in the simulation is introduced. Section 3 shows the simulation results of the radio source size, duration, and position for fundamental waves and harmonic waves at 35\,MHz. The results are discussed and compared in Section 4, and a brief summary is given in Section 5.

\section{Simulation model}

We implemented the three-dimensional (3D) radio wave ray-tracing simulation for a point pulse source on basis of the theory and algorithm proposed by \cite{kontar2019anisotropic}. The simulation solves the Hamilton ray equation and nonlinear Langevin equation corresponding to the Fokker-Planck equation, which can be expressed as
    \begin{align}
        \frac{\mathrm{d} r_i}{\mathrm{d}t} & = \frac{\partial \omega}{\partial k_i} = \frac{c^2}{\omega} k_i,  \label{eq:ki}\\
        \frac{\mathrm{d} k_i}{\mathrm{d} t} & = -\frac{\partial \omega}{\partial r_i} + \frac{\partial D_{ij}}{\partial k_i} + B_{ij} \xi_j,
        \label{eqdk}
    \end{align}
where $r_i\,(i=x,y,z)$ is the position vector of photons, $k_i$ is the wave vector of the radio wave in Cartesian coordinate. $\omega$ is the angular frequency of the wave satisfying the dispersion relation of the unmagnetized plasma: \begin{equation}
    \omega^2=\omega_{pe}^2+c^2k^2.
    \label{eq:dispers}
\end{equation} Here $\omega_{pe}$ is the local plasma frequency expressed as $\omega_{pe}(\bm{r}) = \sqrt{{4\pi e^2 n(\bm{r})}/{m_e}}$ (where $e$ and $m_e$ are respectively the electron charge and mass, and $n$ is the plasma density).
$D_{ij}$ is the diffusion tensor appropriate to scattering, which is given by: 
\begin{equation}
    D_{ij} = \frac{\pi\omega_{pe}^4 }{4\omega c^2}\int q_i q_j S(\bm{q}) \delta(\bm{q}\cdot\bm{k}) \frac{\mathrm{d}^3q}{(2\pi)^3},
    \label{eq:Dij}
\end{equation}
where $\bm{q}$ is the wave-vector of the density fluctuation.  $S(\bm{q})$ is the spectrum of the density fluctuation normalized to the relative density fluctuation variance: 
 $$\epsilon^2= \frac{ \langle \delta n^2\rangle}{ n ^2} = \int S(\bm{q}) \frac{\mathrm{d}^3q}{(2\pi)^3}, $$
where $ n $ is the local average plasma density, taken to be a slowly varying function of position. $B_{ij}$ is a positive-semi-definitive matrix given by $D_{ij} = (1/2) B_{im}B_{jm}^T $. $\bm{\xi}$ is a random vector satisfying the Gaussian distribution with zero mean and unit variance.

To consider the scattering in a medium with anisotropic density fluctuation, an anisotropy tensor 
\begin{equation}
    \mathsf{A} =
  \left( {\begin{array}{ccc}
  1&0&0\\
  0&1&0\\
  0&0&\alpha^{-1}
  \end{array} } \right)
\end{equation}
is introduced. Here  $\alpha$ is the anisotropic parameter representing the ratio of density fluctuation wavelength in the direction of anisotropy and its perpendicular direction. In the model of this work, the density fluctuations are mostly in the direction perpendicular to the solar radius when $\alpha<1$, and 
\textit{vice versa}.   

From Equation \ref{eq:Dij}, one can get the angular scattering rate per unit distance for a classic Gaussian spectrum of density fluctuations \citep{steinberg1971coronal,chrysaphi2018cme,kontar2019anisotropic},
\begin{equation}
    \frac{\mathrm{d}\langle \theta^2\rangle}{\mathrm{d}x} = \frac{\sqrt{\pi}}{2}\frac{\epsilon^2}{h}\frac{\omega^4_{pe}}{(\omega^2-\omega_{pe}^2)^2} 
    \label{eq:angscat}
\end{equation}
where $h$ is the correlation length of density fluctuations.  In this work, the density fluctuation is characterized as power-law distribution with a cutoff at the inner scale of $l_i=(r/R_s)\,[\mathrm{km}]$ and outer scale of $l_o = 0.25 R_s (r/R_s)^{0.82}$ \citep{coles1989propagation,wohlmuth2001radio,krupar2018ItypeIII}, where $R_s$ is the solar radius, \peijin{$r$ is the heliocentric distance, the inner scale and outer scale represent the smallest  and largest wave length of the density fluctuations, respectively}. The angular scattering rate per unit length derived from the power-law fluctuation (Equation (67) in \cite{kontar2019anisotropic}) can be expressed in a similar form of Equation \ref{eq:angscat}  by replacing $h$ with an equivalent scale length $h_{eq}$ given as
\begin{equation}
    h\equiv h_{eq} = l_o^{2/3}l_i^{1/3} /\pi ^{3/2},
    \label{eq:lio}
\end{equation}
and $h$ is a slowly varying function of heliocentric distance $r$. 

For a given wave frequency, the angular scattering rate depends on $\epsilon$, $h$, and the frequency ratio $\omega_{pe}/\omega$.  
As $h^{-1}$ and $\omega_{pe}/\omega$ in Equation \ref{eq:angscat} decreasing with $r$, the scattering rate decreases quickly when the radio waves propagate outward from their original site. 
As a result, the scattering strength experienced by the waves is mainly determined by the coefficient $\eta$ of the angular scattering rate per unit distance in Equation \ref{eq:angscat}, which is defined by
\begin{equation}
    \eta =\epsilon^2/h_0,
\end{equation}
where $h_0=h(r_0)$ is the scale length of density fluctuations near the wave exciting site at $r_0$. In this work, the function $h(r)$ is given, therefore the tuning parameter for scattering rate is the density fluctuation level $\epsilon$ for different simulation cases, but we reiterate that it is $\eta$ the key parameter.


The intensity decay $I/I_0$ due to the  Coulomb collision absorption is described by the optical depth integral along the ray path for each photon:
\begin{align}
            I/I_0 & =  e^{-\tau_a},\\
    \tau_a &= \int \gamma(\bm{r}(t))\, dt,\\
    \gamma &= \frac{4}{3} \sqrt{\frac{2}{\pi}} \frac{e^4 n_e \ln\Lambda}{m v_{Te}^3} \frac{\omega_{pe}^2}{\omega^2},
\end{align}
where $v_{Te}$ is the thermal speed and a constant Coulomb logarithm $\ln\Lambda\simeq20$ is assumed.

\section{Simulation results}

With the anisotropic scattering model, we can perform ray-tracing simulation for the photons launched at a given position. After sufficient steps forward, when the majority of the photons reached 0.95\,AU of heliocentric distance, the final positions ($\bm{r}$) and wave vectors ($\bm{k}$) of the photons near the direction of observation are collected to reconstruct the radio image. The observed position of each photon in the sky plane is estimated by   back-tracking the wave vector. The apparent source intensity map $I(x,y)$ is the weighted photon number density distribution in the sky plane, and the spread of photon arrival times determines the temporal variation of the source.

In the simulation, the background density function $n(r)$ is assumed to be spherically symmetric, the same as the model used in \cite{kontar2019anisotropic}. The source of radio emission is considered to be a point pulse source, the photons are launched at the starting point at the same time. The initiate wave vectors are randomly distributed in an outward direction ($\bm{r}\cdot \bm{k} >0$), namely, a burst source with isotropic emission is considered. 

\begin{figure}[hbt!]
    \centering
    \includegraphics[trim=0.1cm 0cm 0.08cm 0cm, clip=true, width=9.0cm, angle=0]{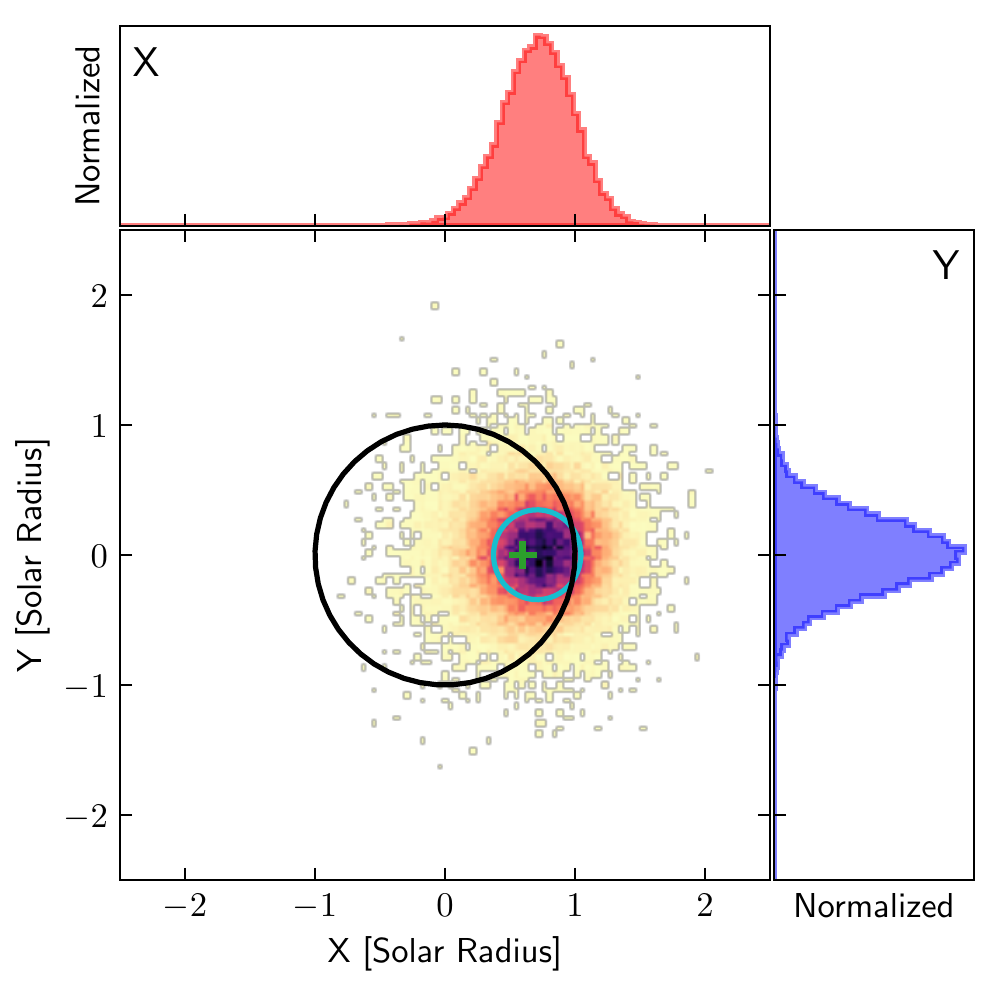}
    \caption{The simulated source intensity image in the sky plane for fundamental emission, which is reconstructed from all collected photons. The frequency of the wave is 35\,MHz with $f_0/f_{pe}=1.1$, $\alpha=0.72$, $\epsilon=0.28$, and $\eta\simeq9\times 10^{-5}\, \textup{km}^{-1}$. A simulation of $3\times10^5$ photons is conducted. The black circle shows the outline of the solar disk, the blue line shows the FWHM of the reconstructed source. The green `+' marks the starting point of the photons. The upper and right panel shows the histogram of the flux intensity in $x$ and $y$ direction.}
    \label{fig:basic}
\end{figure}

\begin{figure}[hbt!]
    \centering 
    \includegraphics[trim=0.1cm 0cm 0.08cm 0cm, clip=true, width=8.5cm, angle=0]{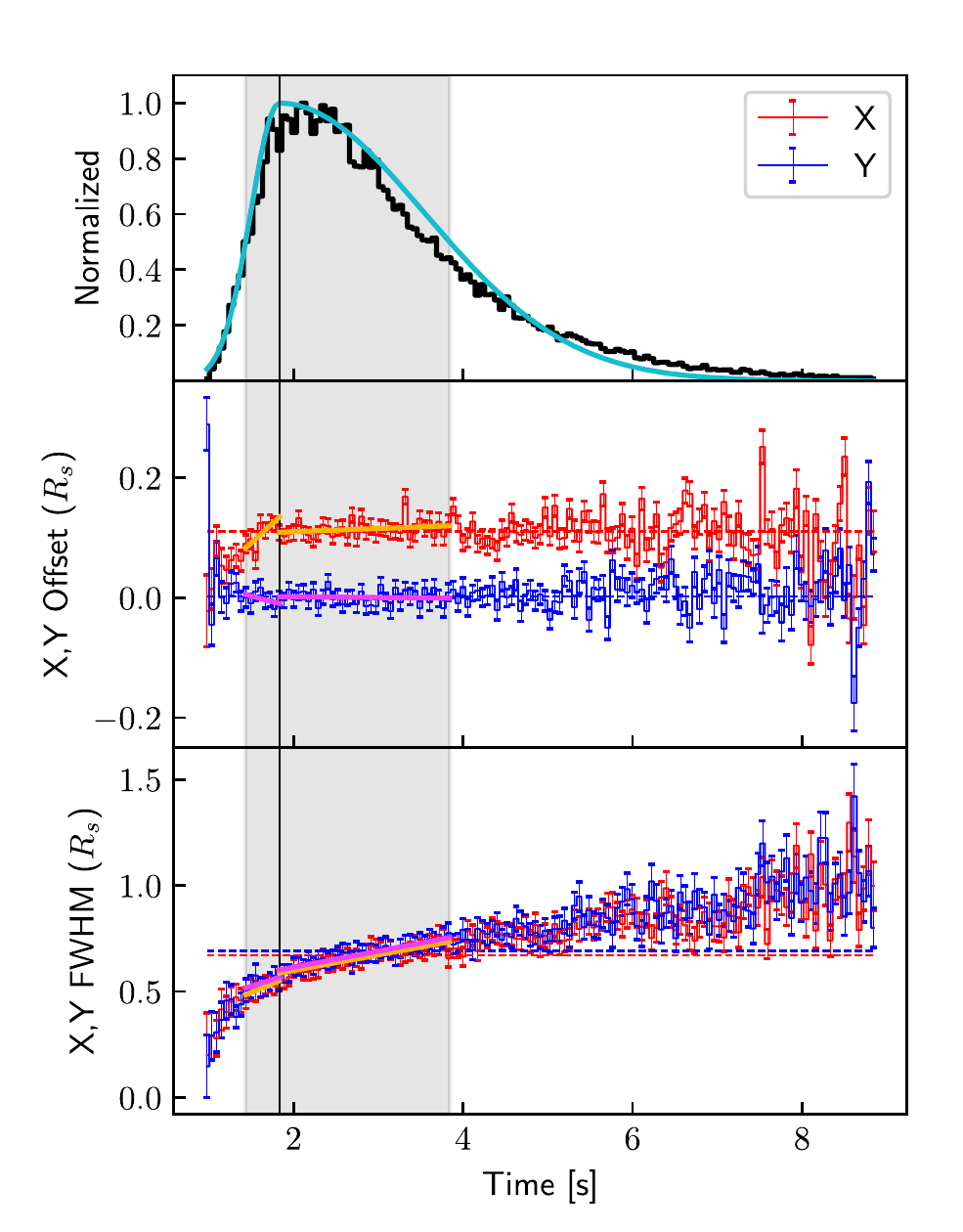}
    \caption{\textit{Upper panel}: the simulated  time-intensity profile. The flux variation is fitted with a double Gaussian function, shown as a cyan line. FWHM range of the fitted flux are marked as gray shadow and the peak of the fitted flux is marked as a black thin line in all panels. \textit{Middle panel}:  the position offset of the simulated source centroid from its original site, where red and blue represents X and Y in heliocentric coordinate. \textit{Lower panel}: the FWHM size of the simulated source. The linear fitting results of position offset and size for the rising and decay phase are shown in the middle and lower panel as \peijin{orange and purple} lines. The simulation parameters are the same as Figure \ref{fig:basic}.} 
    \label{fig:var}
\end{figure}

As a case study, Figure \ref{fig:basic} shows the flux intensity map reconstructed from all collected photons for fundamental emission, representing the time integral imaging of the observed source. 
Figure \ref{fig:var} shows the time-intensity profile and the temporal variations of the source position and size, which is obtained from photons with arrival times within the corresponding time segment. In this simulation case, the radio waves are fundamental waves initiated at the position angle $(\theta_0, \phi_0)=(20^\circ,0^\circ)$, where $\theta$ is the longitude and $\phi$ is the latitude. The density fluctuation level is $\epsilon=0.28$ and the anisotropy parameter is $\alpha=0.72$. {The equivalent correlation length ($h_0$) at the starting point $r_0=1.750R_s$ is about 860\,km, and the corresponding value of the angular scattering rate coefficient is $\eta\simeq9\times 10^{-5}\, \textup{km}^{-1}$ near the wave generation site.}

In Figure \ref{fig:basic}, the source centroid offsets  $0.11 R_s$ in $x$ direction from the starting point (marked as green `+' in the figure). The offset in $y$ direction is negligible. \peijin{This is because the centroid offset is in the radial direction due to the spherical symmetry of the density model used, which is in the $x$ direction in the sky plane for the case shown in this figure. 
} The blue line in Figure \ref{fig:basic} shows the full width half maximum (FWHM) estimation of the reconstructed source. 
The statistical results of time duration, FWHM size, the visual speed, and size expansion rate can be extracted from the reconstructed imaging and its variation.

The FWHM of the time-intensity profile can be divided into two phases, namely the rising and decay phase. We fitted the flux variation with a double Gaussian function \citep{zhang2019source}:
\begin{equation}
	G(t) = A \exp{\left(-\frac{(t-t_0)^2}{ \tau^2}\right)}, \qquad \tau = \left\{ \begin{aligned}
	\tau_{R}, & & {t \leq t_0}\\
	\tau_{D}, & & {t >  t_0}
	\end{aligned}\right. .
	\end{equation}
The fitted result is shown as cyan line in Figure \ref{fig:var}. The peak and the FWHM of the flux is obtained from the fitted curve. 
In Figure \ref{fig:var}, the FWHM range marked by gray shadow is divided by a thin vertical line at the peak of flux ($t=1.83$\,s).
The regime before the peak is the rising phase, and after is the decay phase. The duration of the rising phase and decay phase are $\tau_R=0.4$ and $\tau_D=2.0$ seconds for this case. 

The source offset measures the vector distance between the observed centroid and the starting point of the photons \peijin{in the sky plane}. In this case, the average offset is $0.105 R_s$ during the rising phase and $0.115R_s$ during the decay phase, the offset in y direction is negligible (less than $0.002R_s$). 

The source size is measured by the FWHM in $x$ and $y$ direction: $\textup{FWHM}_{x,y} = 2\sqrt{2\ln{2}}\sigma_{x,y}$,
where $\sigma_{x,y}$ is the variance of the distribution. In this case, the average FWHM in $x$ direction is $0.513R_s$ $(0.137^\circ)$ for the rising phase and $0.662R_s$ $(0.176^\circ)$ for the decay phase. The average FWHM in $y$ direction is $0.538R_s$ $(0.143^\circ)$ for the rising phase and $0.680R_s$ $(0.181^\circ)$ for the decay phase. 
The source FWHM size in y direction is slightly larger than that in $x$ direction.

The speed and expansion rate of the source in the sky plane can be obtained from the linearly fit of the variation of the source position and size with time.
From Figure \ref{fig:var}, we can see that, the source behaves differently in movement and expansion during the rising and decay phase, for this case, the moving speed and the expansion rate are both larger in the rising phase.
The source size has a positive expansion rate in both $x$ and $y$ direction.

In order to study the properties and variations of the source in different parameter sets, namely the density fluctuation level and the anisotropic scale, we run the above simulation process in the parameter space of $\alpha\in[0.05, 0.99]$ and $\epsilon\in[0.03, 0.45]$. We uniformly select $36 \times 36$ points in the parameter space. For each case of parameter set  ($\epsilon,\alpha$), a simulation of $3\times10^5$ photons is conducted. 
In each case, the frequency of the wave is assumed to be 35\,MHz, the frequency ratio between the wave and local plasma frequency at the starting point ($f_0/f_{pe}$) is 1.1 for fundamental emission and 2.0 for harmonic emission. The photons are launched at the center of the solar disk ($\theta_0=\phi_0=0$), for fundamental emission $r_0=1.750R_s$ and $h_0=860$\,km, for harmonic emission $r_0=2.104R_s$ and $h_0=1010$\,km. The angular scattering rate coefficient $\eta$ changes in the range between $8.9\times 10^{-7} \, \textup{km}^{-1}$ to $2.4\times 10^{-4}\, \textup{km}^{-1}$ for $\epsilon\in[0.03, 0.45]$.
The  statistical result of the source properties is presented in the following subsections.

\subsection{Source size and duration}
\label{sec:sizepos}

The source size and duration are the most common information that can be extracted from the imaging and spectroscopy observations of solar radio bursts. In this subsection, the source size is measured as the intensity weighted FWHM size for all collected photons in the simulation as done in Figure \ref{fig:basic}. The decay time is measured as the time of intensity decreasing by $1/e$ from the peak, namely, $\tau_D$.

\begin{figure}[hbt!]
	\centering
	\includegraphics[trim=0.1cm 0cm 0.38cm 0cm, clip=true, width=7.7cm, angle=0]{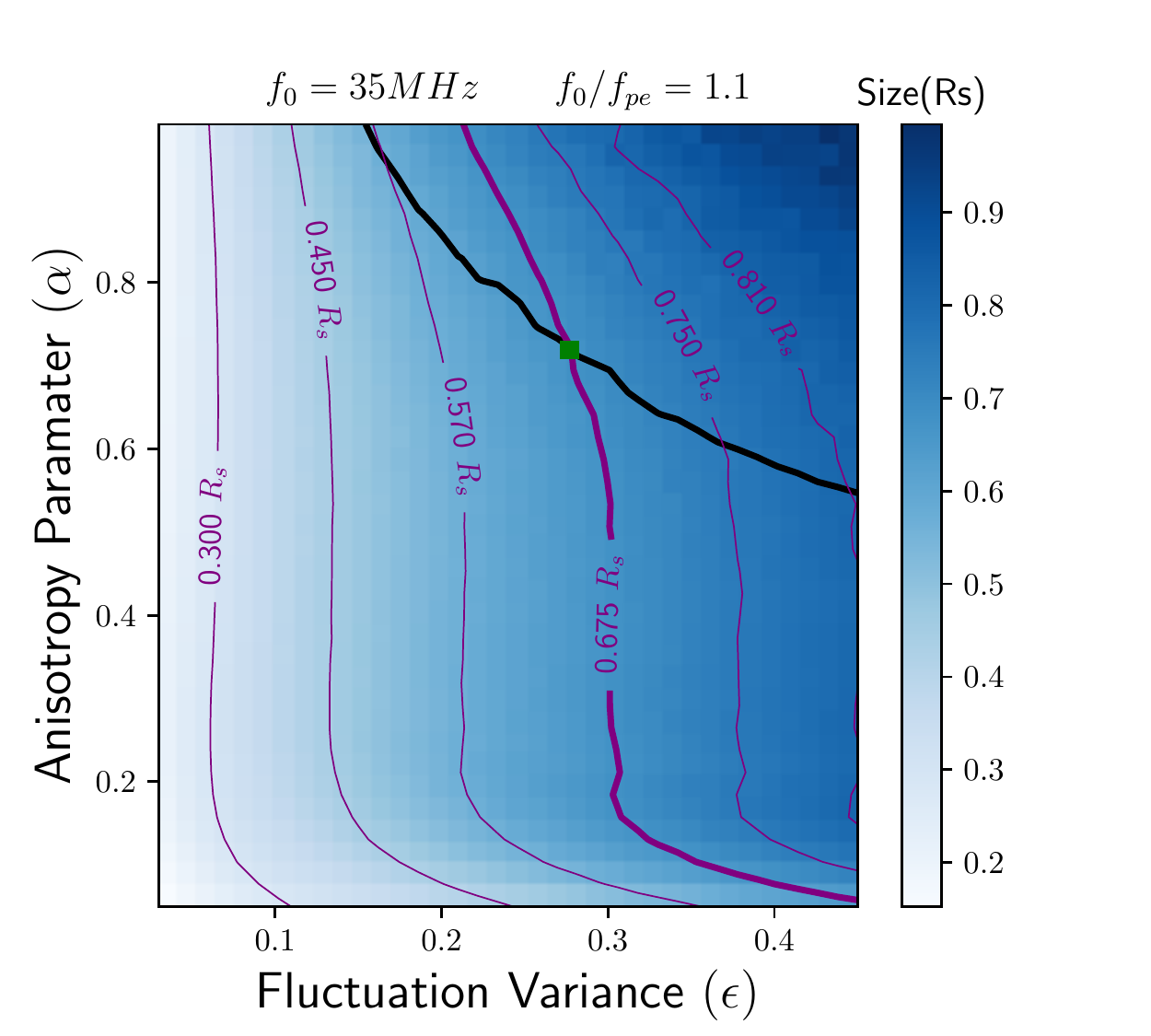}
	\includegraphics[trim=0.1cm 0cm 0.38cm 0cm, clip=true, width=7.7cm, angle=0]{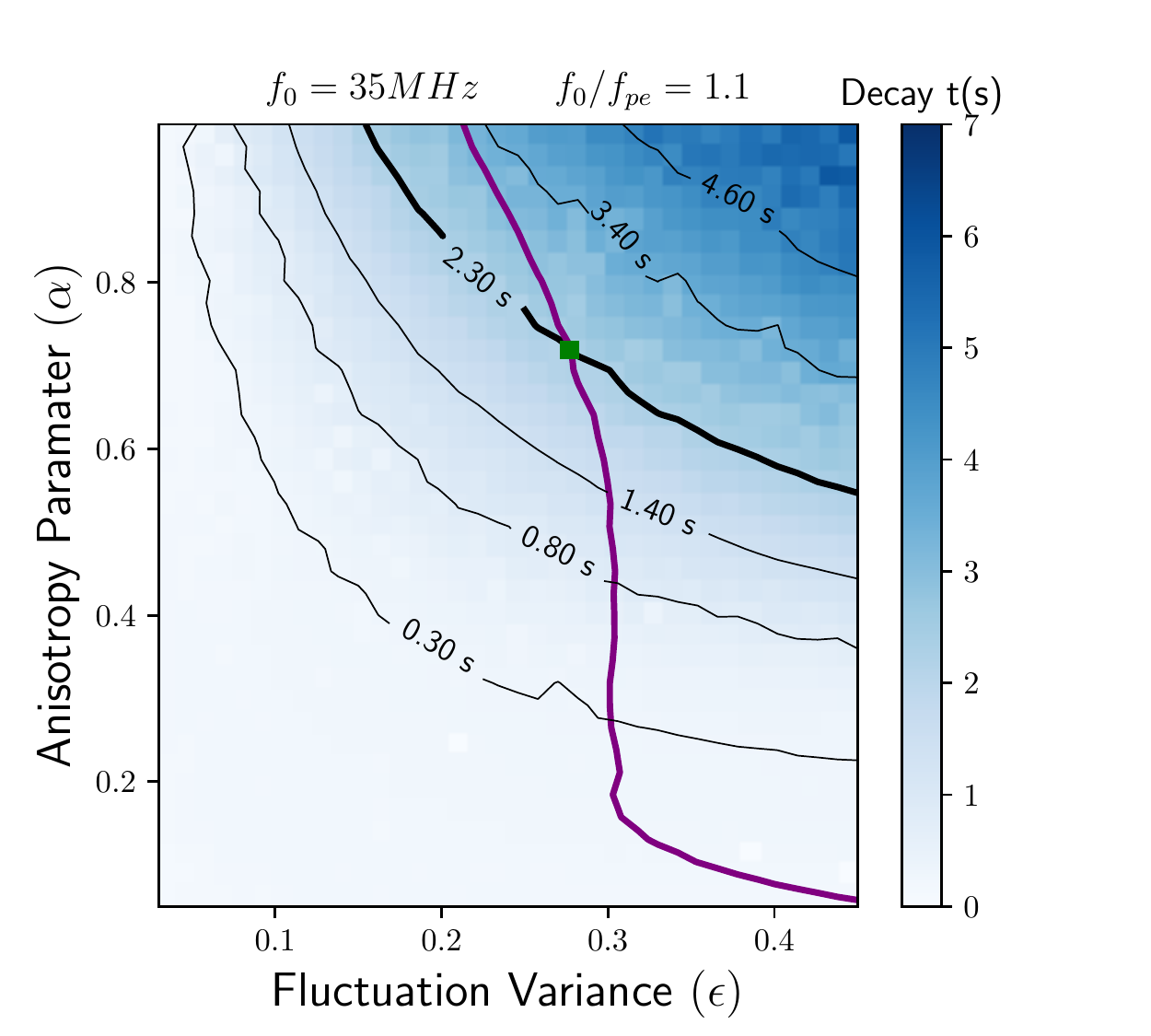}
	\caption{The source size and decay time of the simulation results for the fundamental emission. Purple lines in the left panel show the contour of source size, and black lines in the right panel show the contour of the decay time for the simulation data. The bold lines mark the value of $D_{\rm{FWHM}}=0.675 R_s$, $\tau_D=2.30\, \rm{s}$ for $f_0=35$\,MHz in Equation \ref{eq:D} and \ref{eq:tau}. The green square marks the cross point of the bold contour lines at $\epsilon=0.277$ and $\alpha = 0.719$. The equivalent scale length of density fluctuation $h_0$ is about 860\,km near the wave generation site. 
	\label{fig:sizet01}}
\end{figure}

The source size and decay time of the fundamental emission ($f_0/f_{pe}=1.1$) is shown in Figure \ref{fig:sizet01}. 
Within the parameter space, the source size varies from 0.2 to 0.95 solar radii and the decay time varies from about 0 to 7 seconds. 
The source size is largely determined by the fluctuation variance value (or the scattering coefficient $\eta$), the tendency of contour lines in Figure \ref{fig:sizet01}a is nearly parallel to y-axes in the range of $\alpha>0.1$. The decay time is sensitive to both $\alpha$ and $\epsilon$ (or $\eta$). The decay time increases with the density fluctuation level and the anisotropy parameter, thus decrease with the anisotropy degree in the density fluctuation.

\begin{figure}[hbt!]
	\centering
	\includegraphics[trim=0.1cm 0cm 0.38cm 0cm, clip=true, width=7.7cm, angle=0]{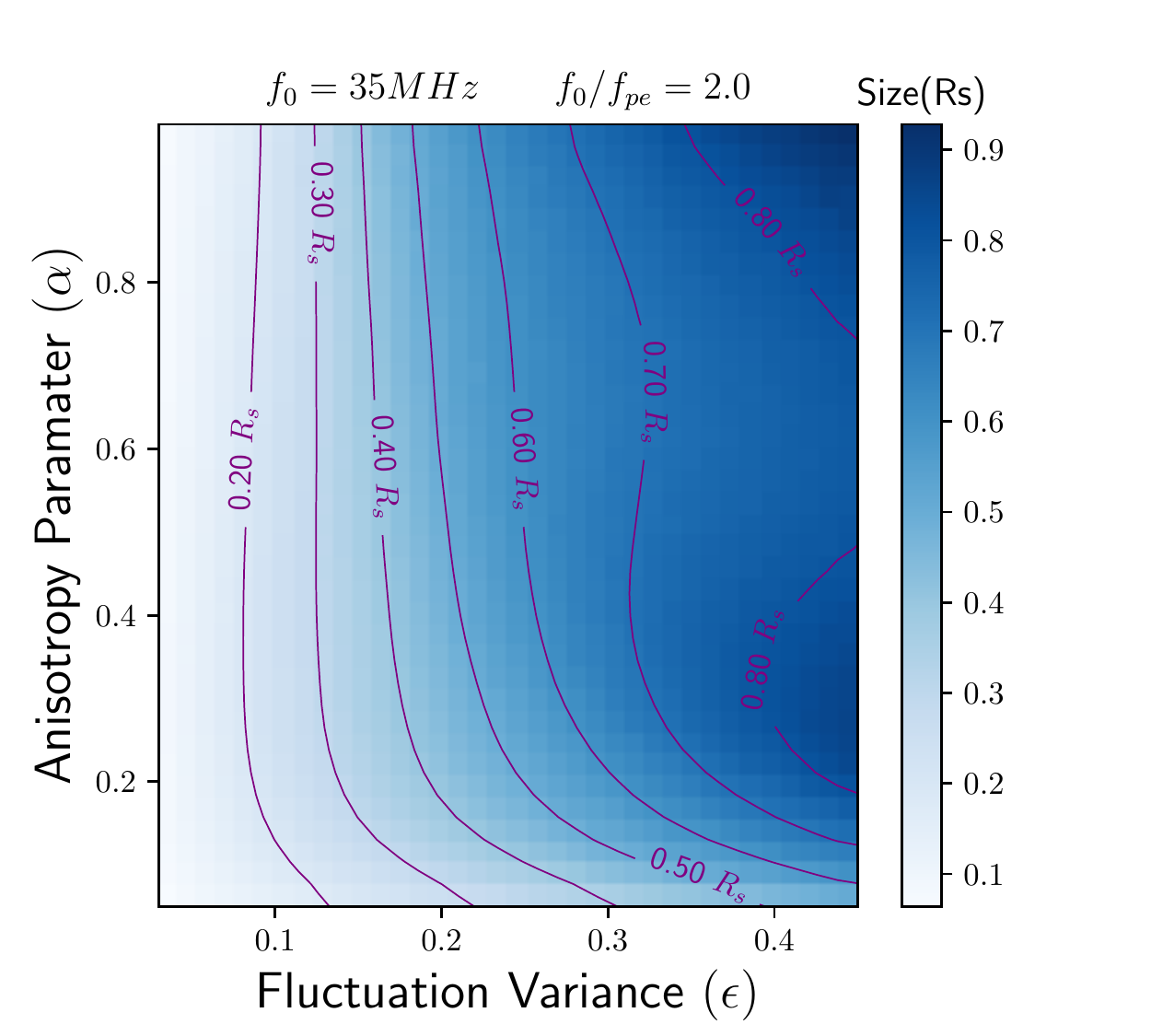}
	\includegraphics[trim=0.1cm 0cm 0.38cm 0cm, clip=true, width=7.7cm, angle=0]{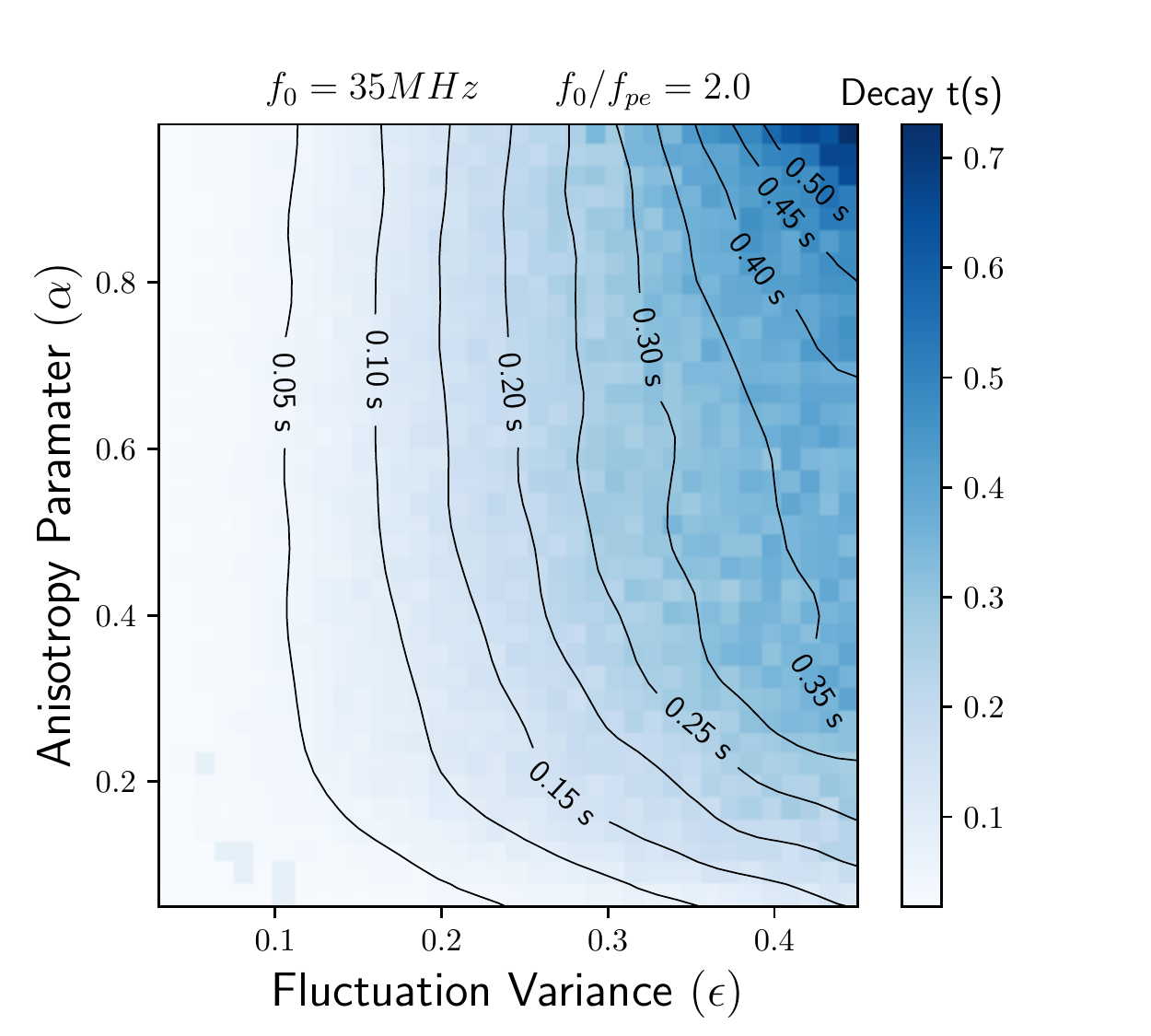}
	\caption{The source size and decay time of the simulation results for the harmonic emission, the specs are the same as Figure \ref{fig:sizet01}, and   $h_0=1010$\,km.
	\label{fig:sizet02}}
\end{figure}

For the harmonic emission with the starting frequency of $f_0/f_{pe}=2.0$, as shown in Figure \ref{fig:sizet02} the source size varies from 0.1 to 0.9 $R_s$ and the decay time varies from 0 to 0.7 seconds within the parameter space. Compared to the fundamental emission with $f_0/f_{pe}=1.1$, the source size of harmonic emission is slightly smaller than the size of fundamental emission with a value of about 0.1\,$R_s$, while the decay time of harmonic emission is significantly smaller than the fundamental emission. Both the source size and duration are not sensitive to the variation of fluctuation anisotropy for $\alpha>0.2$.

The simulation results in the parameter space can be used to constrain the background plasma properties, compared with the corresponding observation. According to the previous fitted result of multiple observations for the source size and decay time \peijin{of type III radio bursts} \citep{kontar2019anisotropic}, the relationship between the observed source size [Degree] and frequency [MHz] can be expressed as:
\begin{equation}
\label{eq:D}
D_{\rm{FWHM}} =  (11.78\pm 0.06) \times f^{-0.98\pm0.05}.    
\end{equation}
The decay time [s] and frequency [MHz] can be expressed as:
\begin{equation}
\tau = (72.23\pm0.05) \times f^{-0.97\pm0.03}.    
\label{eq:tau}
\end{equation}
The observed source size and decay time of the 35\,MHz source are $0.18^\circ$ (0.675\,$R_s$) and 2.30 seconds accordingly. Figure \ref{fig:sizet01} shows the simulation results of the source size and decay time with different density fluctuation variance and the anisotropy parameter. The parameter set $\epsilon=0.277$ and $\alpha = 0.719$ (marked as a green square in Figure \ref{fig:sizet01}) can satisfy both the observation estimation of size and decay time at 35\,MHz. 
The corresponding scattering rate coefficient for $\epsilon=0.277$ in the simulation is $\eta = 8.9\times 10^{-5}\, \textup{km}^{-1} $.
The source properties variations of $\epsilon=0.28$ and $\alpha = 0.72$ is shown in Figure \ref{fig:basic} and \ref{fig:var}, note that the starting point of the photons is \peijin{chosen to be located at $\theta_0=20^\circ$ in Figure \ref{fig:basic} and \ref{fig:var} for displaying the source position offset}.
 
\subsection{Position offset of the source}

In Section \ref{sec:sizepos}, the starting point is set as in the center of the solar disk, and the centroid of the reconstructed source is located near $(0,0)$. Thus, the offset of the source from the starting point is close to zero when the position angle of the starting point $\theta_0$ is 0. When the position angle of the starting point is away from zero, the offset is not negligible. 
Considering the sphere-symmetric of the background parameters, the simulated observation for a given $\theta_0$ can be obtained by rotating the result wave vector ($\bm{k}$) and ($\bm{r}$) of all photons to  the reverse angle direction of $\theta_0$ and redo the photon collecting and image reconstruction process. 
For simplification without losing generality, we use $y_0=0$ for the starting point \peijin{so that the simulated offset is mainly in the $x$ direction}.
\peijin{We measure} the offset in the $x$ direction as $\Delta x= x_c^{obs}-x_0$. The source centroid $x_c^{obs}$ is measured as the weighted average position in $x$ axis for all collected photons as in Figure \ref{fig:basic}. 

\begin{figure}[hbt!]
	\centering
	\includegraphics[trim=0.0cm 0cm 0.0cm 0cm, clip=true, width=7.7cm, angle=0]{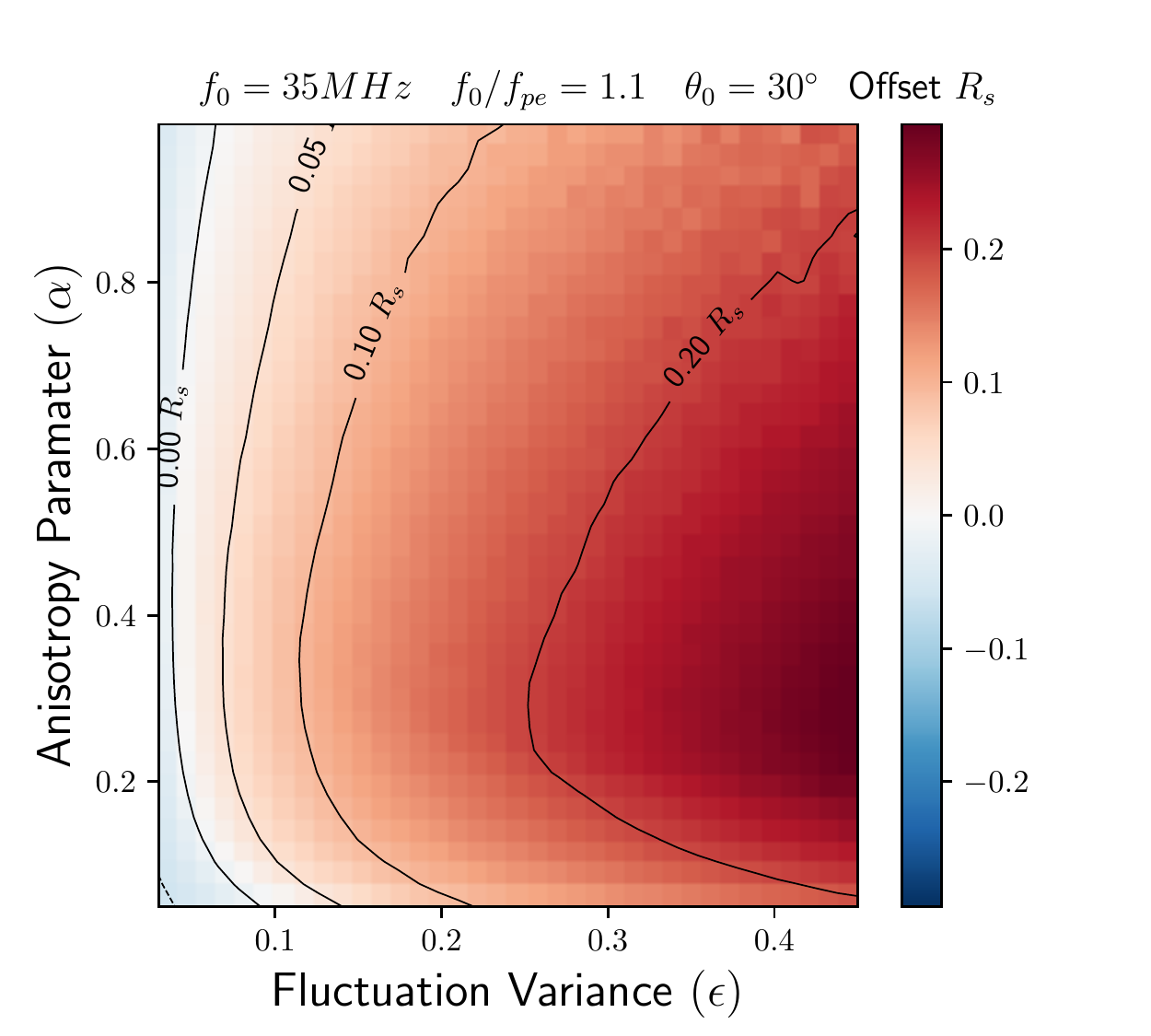}
	\includegraphics[trim=0.0cm 0cm 0.0cm 0cm, clip=true, width=7.7cm, angle=0]{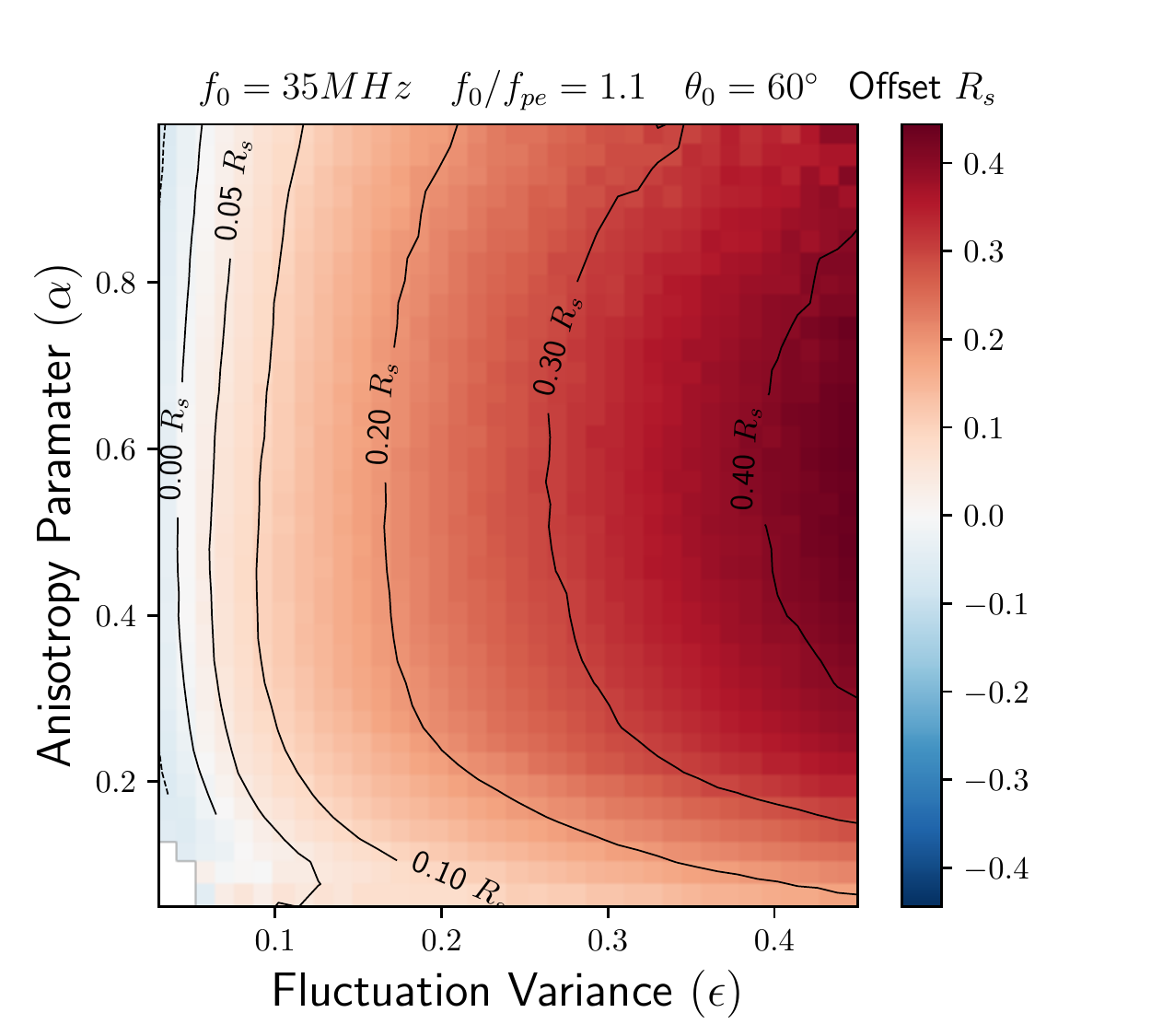}	
	\caption{The offset of the reconstructed source centroid from the starting point of the fundamental emission. The left and right panels show the result of the source locations at $\theta_0=30^\circ$ and $60^\circ$, respectively. The positive value represents the outward offset from the solar disk center, the negative value represents the inward offset in this figure. The equivalent scale length of density fluctuation $h_0$ is about 860\,km near the wave generation site.}
	\label{fig:offset01}
\end{figure}

\begin{figure}[hbt!]
	\centering
	\includegraphics[trim=0.0cm 0cm 0.0cm 0cm, clip=true, width=7.7cm, angle=0]{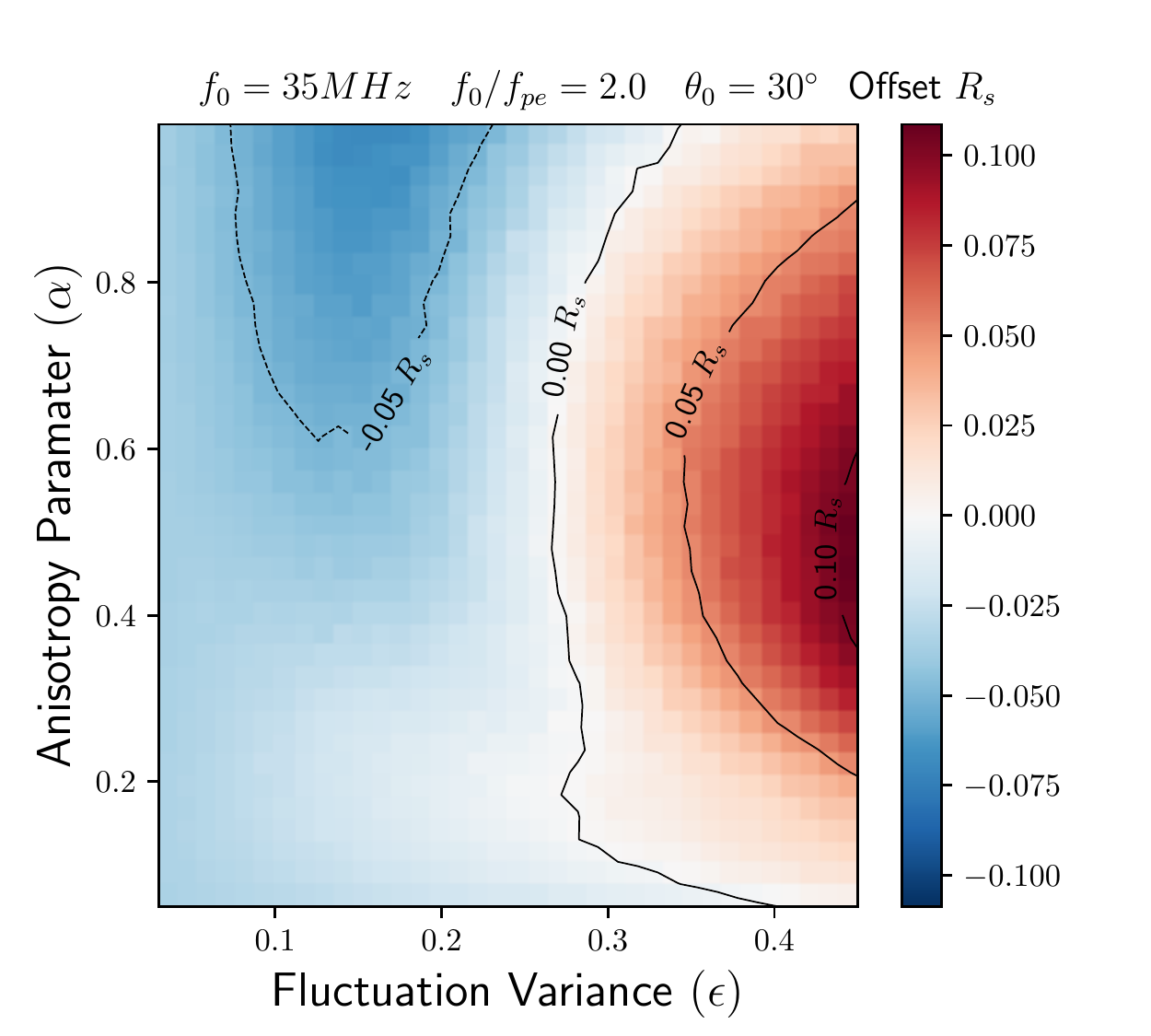}
	\includegraphics[trim=0.0cm 0cm 0.0cm 0cm, clip=true, width=7.7cm, angle=0]{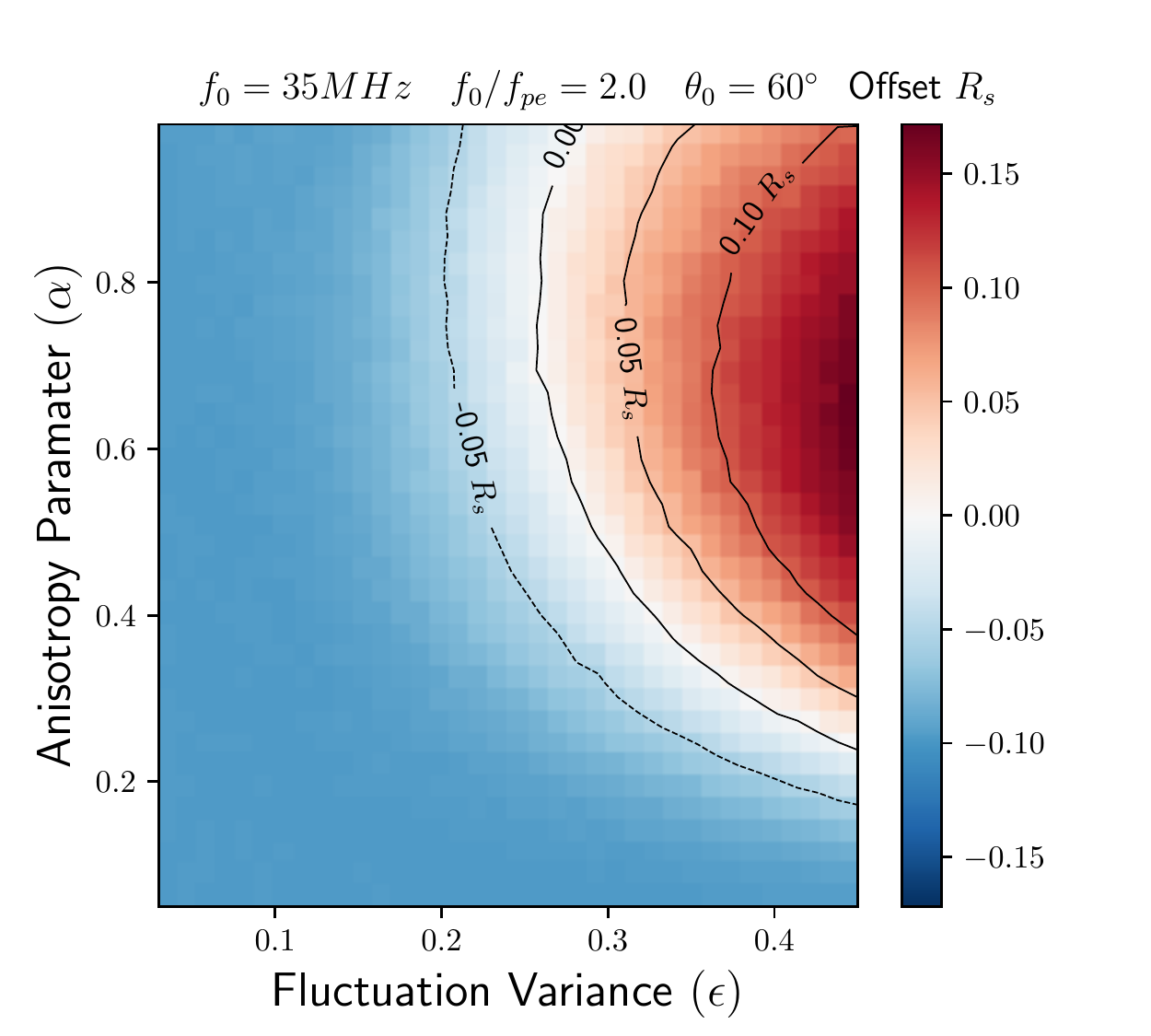}
	\caption{The offset from the reconstructed source centroid and the starting point of the photons for harmonic emission. The specs are the same as Figure \ref{fig:offset01}, and   $h_0=1010$\,km.}
	\label{fig:offset02}
\end{figure}

Different density fluctuation variance and anisotropic scale can result in different offset amounts. Figure \ref{fig:offset01} and \ref{fig:offset02}  shows the source offset ($\Delta x$) of the fundamental ($f_0/f_{pe}=1.1$) and harmonic ($f_0/f_{pe}=2.0$) emission in the parameter space  of $\epsilon\in[0.03, 0.45]$ and  $\alpha\in[0.05, 0.99]$. 
For the fundamental emission, as shown in Figure \ref{fig:offset01}, the result of the offset value is positive for a major part of the parameter space, meaning the direction of the offset is outward. In a small portion of the parameter space at $\epsilon<0.03$ with weak scattering, the offset is inward. The source offset $\Delta x$ increases with the density fluctuation variance  $\epsilon$ for the two starting position angles $\theta_0=30^\circ$ and  $\theta_0=60^\circ$. The offset of the source with a starting position angle of $60^\circ$ is larger than  that of $30^\circ$. The maximum offset of $\theta_0=30^\circ$ is about $0.28\,R_s$ and the maximum offset of $\theta_0=60^\circ$ is about $0.43\,R_s$. 

The source offset of harmonic emission is largely different from the fundamental emission, as shown in  Figure \ref{fig:offset02}. Considerable portions of the results in parameter space have inward offset with $\Delta x<0$. The outward offset can exist when the density fluctuation level is high and the anisotropy degree is low, for example, $\epsilon>0.3$ and $\alpha>0.3$. The offset value of $\theta_0=30^\circ$ varies in $(-0.07,0.11)$ within the parameter space. The offset value of $\theta_0=60^\circ$ varies in $(-0.10,0.17)$ within the parameter space. 
Comparing the offset of the fundamental and harmonic. The offset value of the harmonic emission is much smaller than that of the fundamental emission.

\begin{figure}[hbt!]
	\centering
	\includegraphics[trim=0.0cm 0cm 0.0cm 0cm, clip=true, width=8.cm, angle=0]{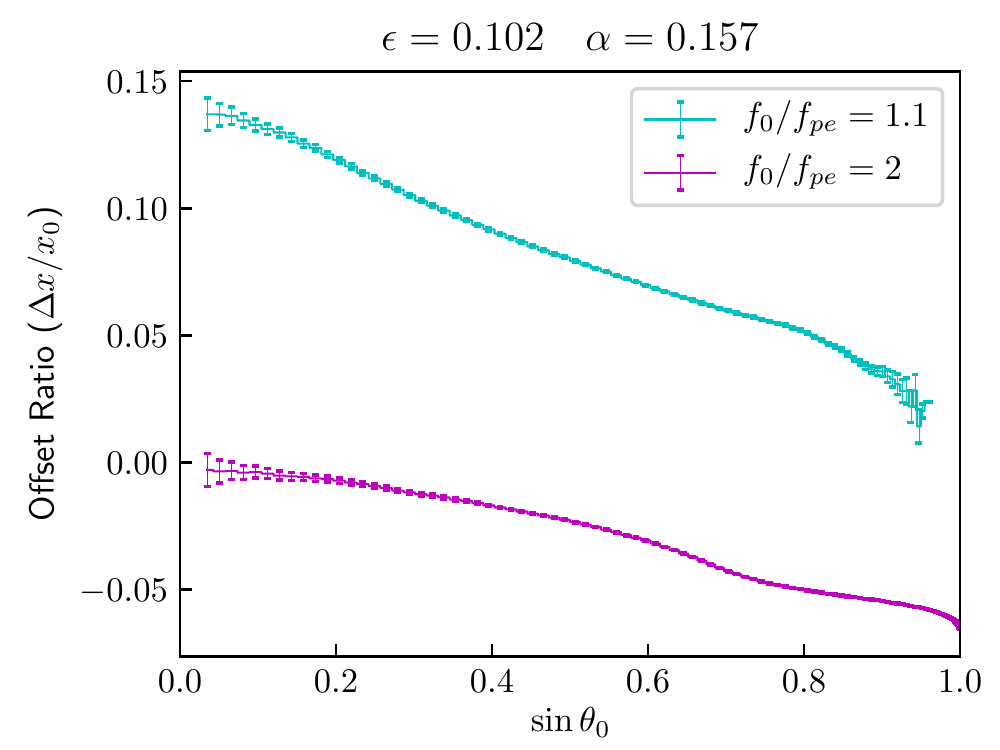}
	\includegraphics[trim=0.0cm 0cm 0.0cm 0cm, clip=true, width=8.cm, angle=0]{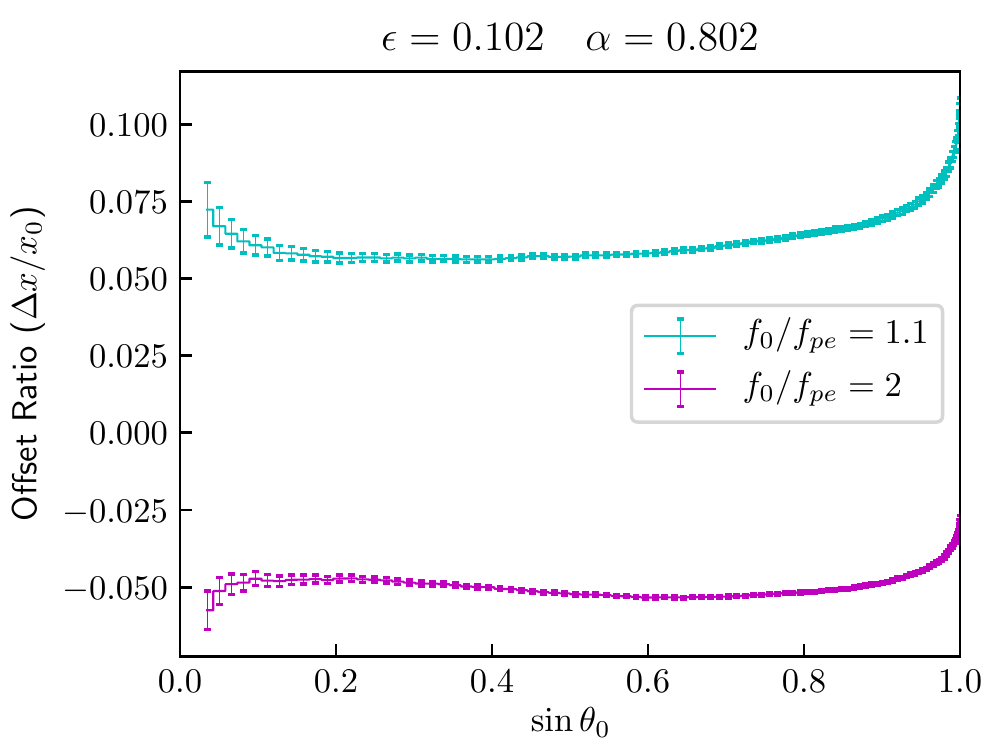}	 
	
	\includegraphics[trim=0.0cm 0cm 0.0cm 0cm, clip=true, width=8.cm, angle=0]{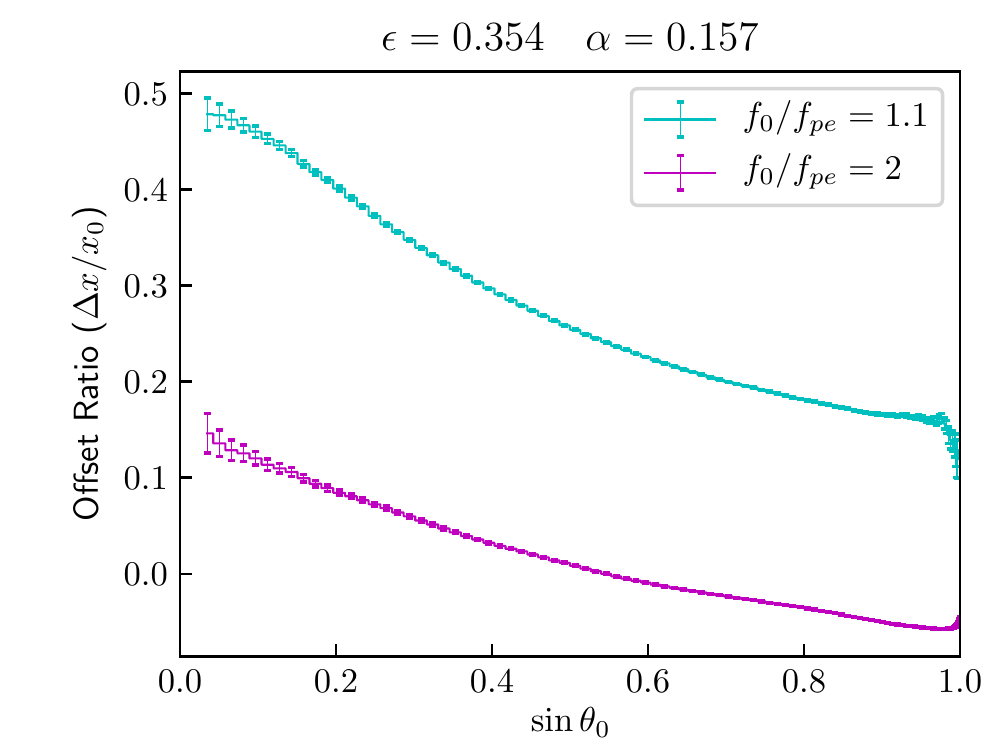}
	\includegraphics[trim=0.0cm 0cm 0.0cm 0cm, clip=true, width=8.cm, angle=0]{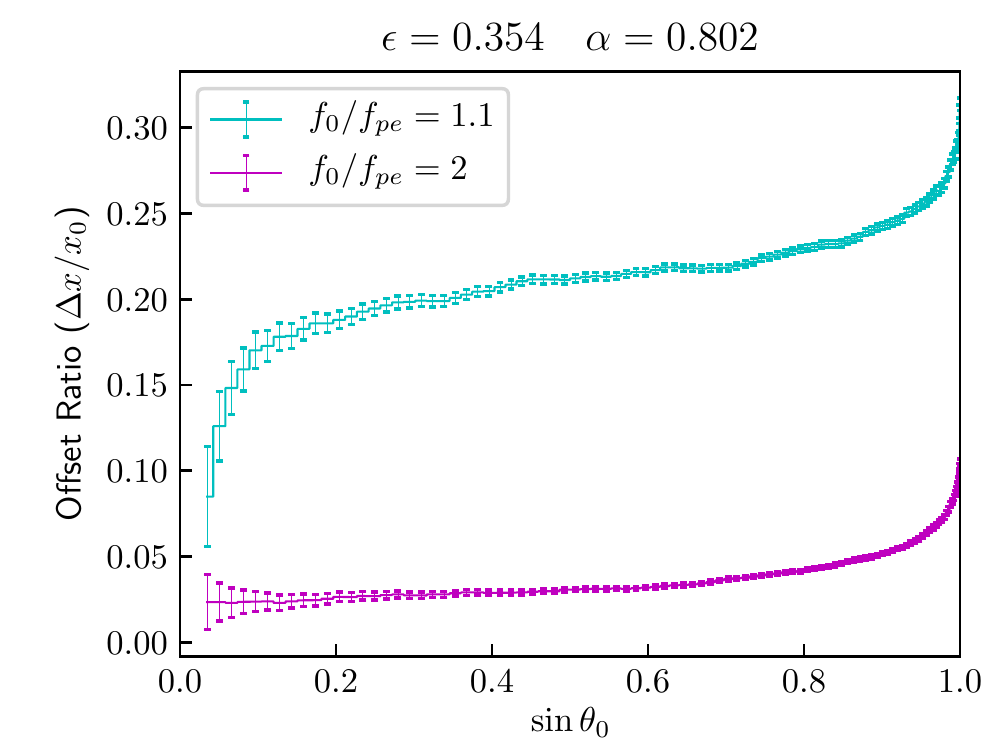} 
	
	\caption{The offset ratio ($\Delta x/x_0$) of the source centroid from the starting point of the photons with different starting position angle. The simulation parameter sets are labeled on the top of each panel. \label{fig:offset_ang_ratio}}
\end{figure}

For further study of the source offset, we selected four cases with the parameter of $\epsilon=0.102, 0.354$, and $\alpha=0.157, 0.802$. These four cases represent the relatively high and low value of angular scattering rate and a relatively large and small degree of anisotropy.
Figure \ref{fig:offset_ang_ratio} shows the offset ratio ($\Delta x/x_0$) of the source centroid to the starting point. The four panels of Figure \ref{fig:offset_ang_ratio} show that for all these four parameter sets, the offset of fundamental emission is more outward than the harmonic emission.
For relatively small anisotropy parameter $\alpha$ (high degree of anisotropy), the relative offset for both fundamental and harmonic emission decrease with the position angle $\theta_0$.  For relatively larger anisotropy parameter $\alpha$ (low degree of anisotropy), the relative offset is stable when $\sin{\theta_0}<0.8$, and increase with position angle when $\sin{\theta_0}>0.8$. Comparing the upper and lower column of Figure \ref{fig:offset_ang_ratio}, one can find that the value of relative offset is larger when the density fluctuation (or angular scattering rate) is larger.

	
	
	

\begin{figure}[hbt!]
	\centering
	\includegraphics[trim=0.0cm 0cm 0.0cm 0cm, clip=true, width=8.cm, angle=0]{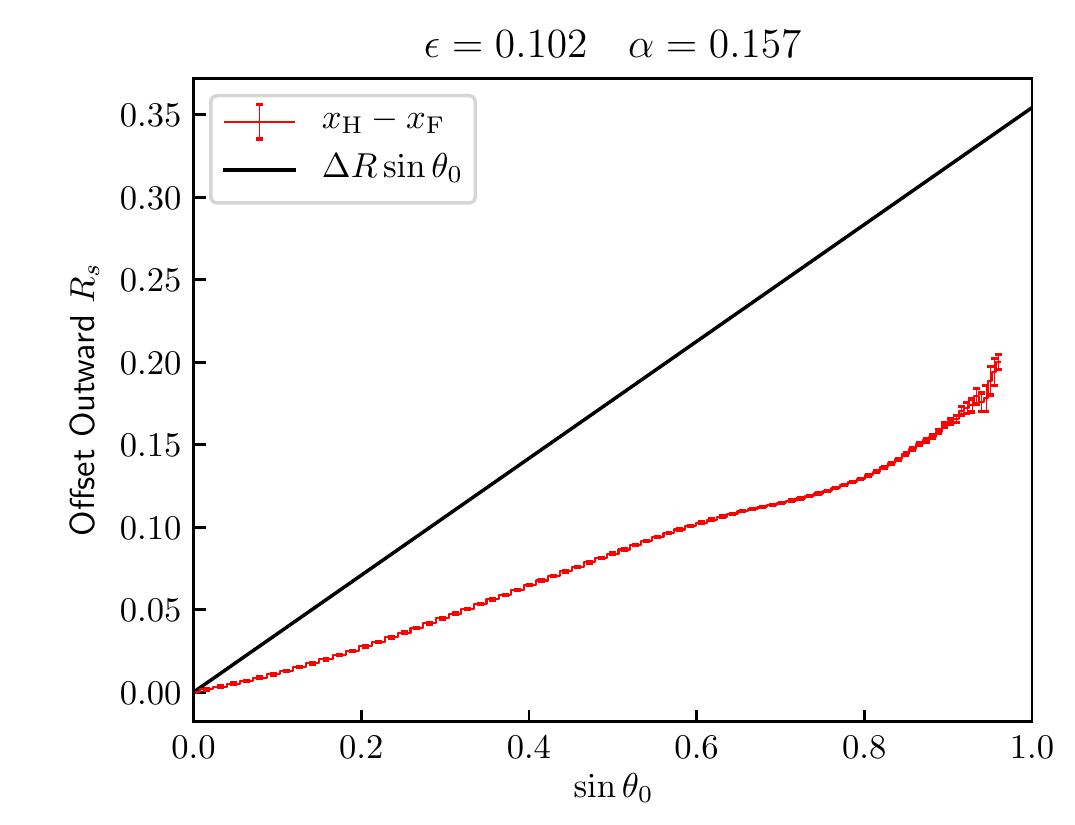}
	\includegraphics[trim=0.0cm 0cm 0.0cm 0cm, clip=true, width=8.cm, angle=0]{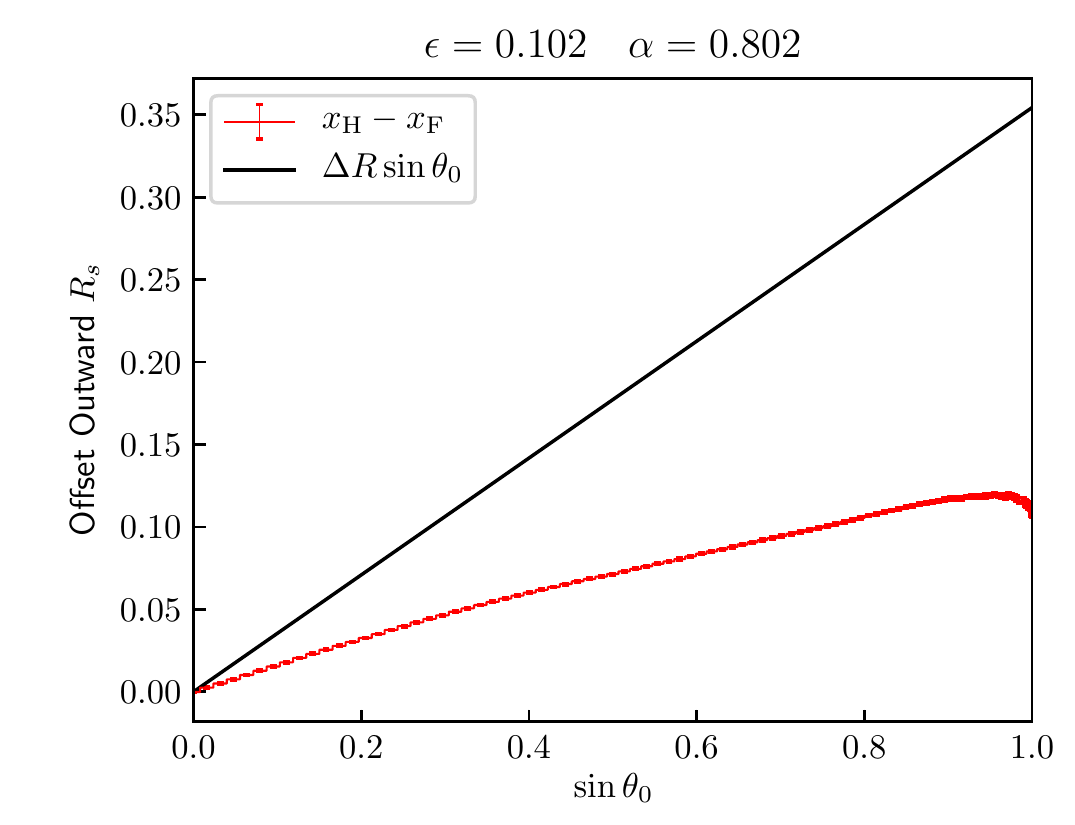}	 
	
	\includegraphics[trim=0.0cm 0cm 0.0cm 0cm, clip=true, width=8.cm, angle=0]{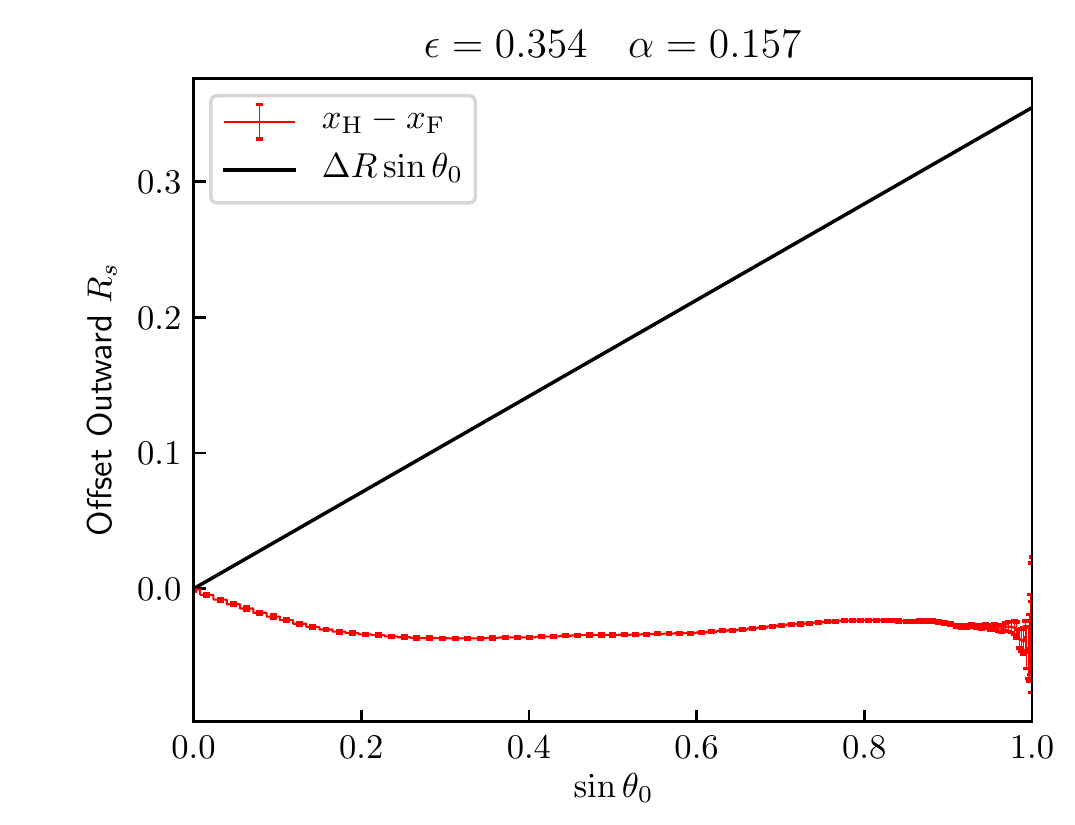}
	\includegraphics[trim=0.0cm 0cm 0.0cm 0cm, clip=true, width=8.cm, angle=0]{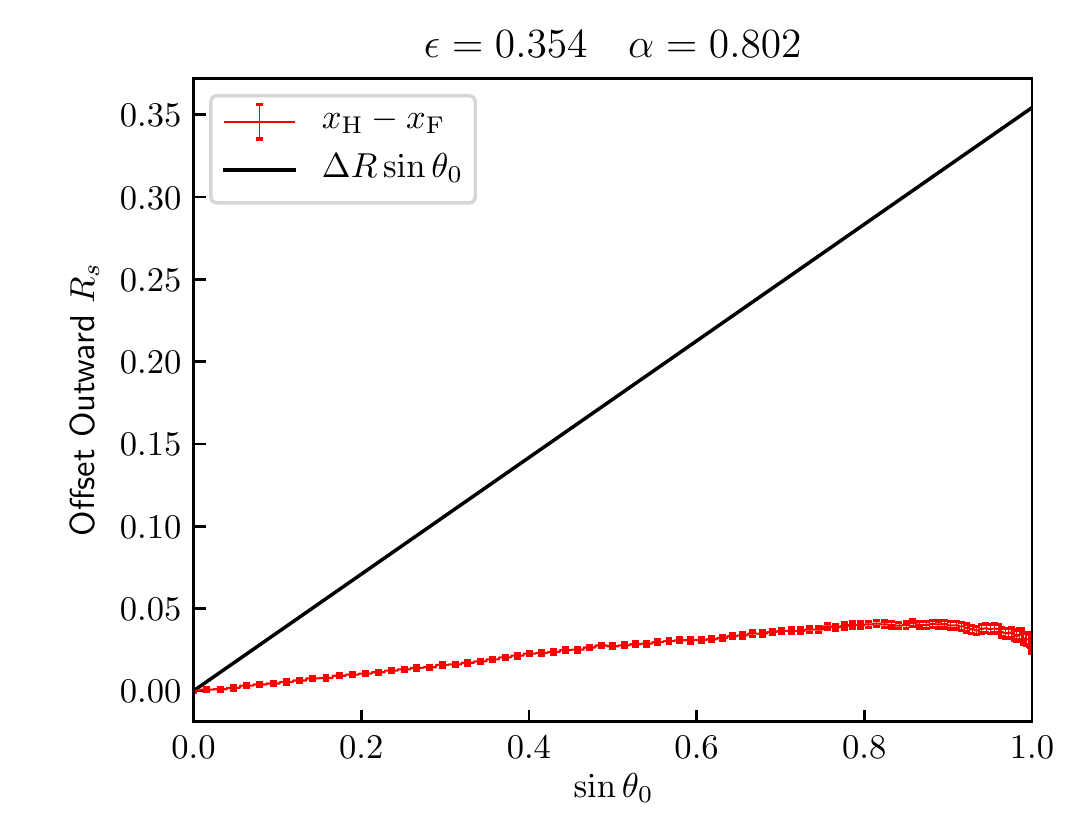} 
	
	\caption{The distance of the reconstructed source centroid of fundamental and harmonic emission (red lines), compared with the geometry distance of the starting point in the sky plane (black lines). \label{fig:offset_dist}}
	
\end{figure}

Assuming the wave of fundamental and harmonic emission with the same given frequency is generated at the same angular direction from the solar center. The visual distance between the starting points of the fundamental and the harmonic is $\Delta R \sin\theta_0$ in the sky plane, where $\Delta R$ is the height difference between the starting points of the fundamental and harmonic wave. The source offset due to the refraction and scattering will cause the observed visual distance to deviate from $\Delta R \sin\theta_0$. 
In this work, for the wave of 35\,MHz, the fundamental wave is generated at $R_F=1.750R_s$ by assuming $f_0/f_{pe}=1.1$, the harmonic wave is generated at $ R_H=2.104R_s$ by assuming $f_0/f_{pe}=2.0$, $\Delta R$ for this case is $0.354 R_s$. 
Figure \ref{fig:offset_dist} shows the distance between the reconstructed source centroids of fundamental and harmonic for 35\,MHz. From Figure \ref{fig:offset_dist} we can see that the value of the distance between the reconstructed sources is significantly smaller than the visual distance of the starting points of the two waves. In other words, the apparent sources of the fundamental and harmonic emission will be much closer with each other than their true source in the sky plane.

\FloatBarrier
The influence of the background medium to the visual source position can be divided into two parts, namely the refraction and scattering, marked as `refr' and `scat' in the following discussion. 
With the dispersion relation of unmagnetized cold plasma described in Equation (\ref{eq:dispers}), the motion direction of the photon is aligned with the wave vector as shown in Equation \ref{eq:ki}. Thus, the variation of the wave vector leads to the bending of the ray-path of the photon, and eventually causes the offset of the visual position of the source centroid.
The variation of the wave vector in Equation \ref{eqdk} can be split into these two parts accordingly:
\begin{align}
    \frac{\mathrm{d} k_{i,\mathrm{refr}}}{\mathrm{d} t} &=  -\frac{\partial \omega}{\partial r_i},\\
    \frac{\mathrm{d} k_{i,\mathrm{scat}}}{\mathrm{d} t} & =\frac{\partial D_{ij}}{\partial k_i} + B_{ij} \xi_j.
\end{align}
In the ray tracing process, 
we can measure the cumulative change of the wave vector $k_i$ due to the refraction and scattering by individually integrating the variation of wave vector due to these two factors for each photon:
\begin{align}
    \Delta K_{i,\mathrm{refr}} &=  \sum \mathrm{\delta}k_{i,\mathrm{refr}},\\
    \Delta K_{i,\mathrm{scat}} & =\sum \mathrm{\delta}k_{i,\mathrm{scat}}.
\end{align} 

\begin{figure}[hbt!]
	\centering
	\includegraphics[trim=0.0cm 0cm 0.0cm 0cm, clip=true, width=8.4cm, angle=0]{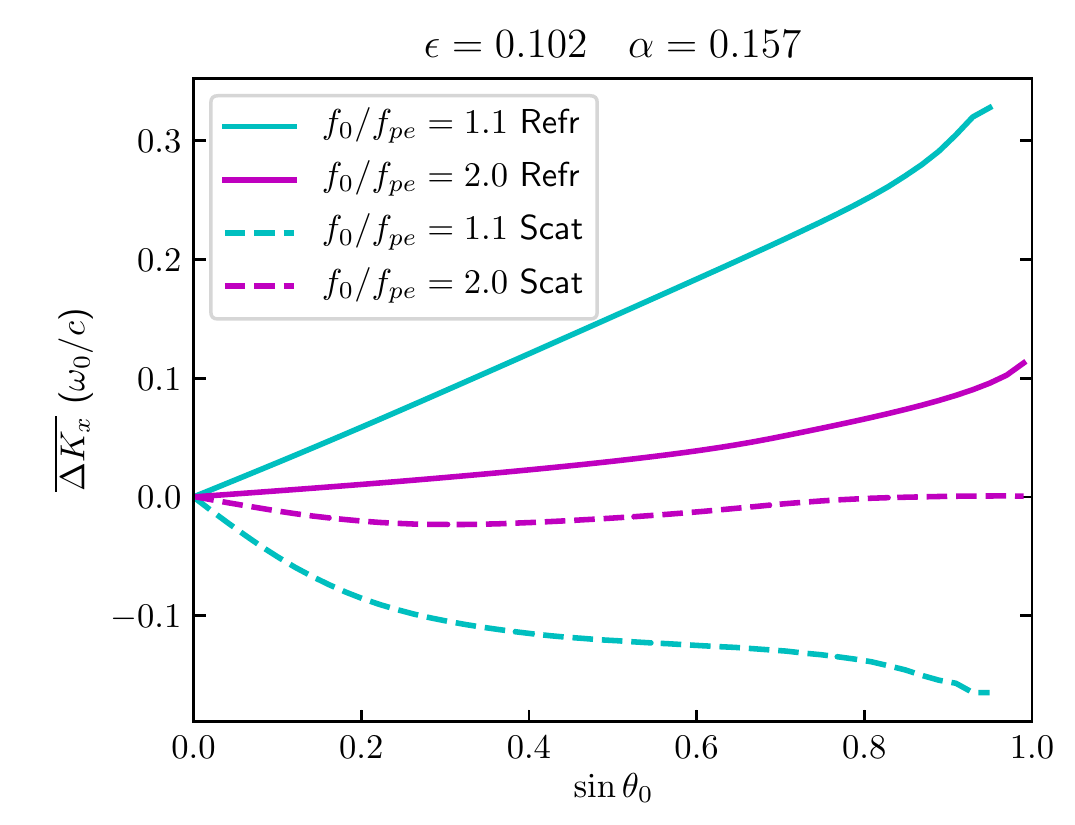}
	\includegraphics[trim=0.0cm 0cm 0.0cm 0cm, clip=true, width=8.4cm, angle=0]{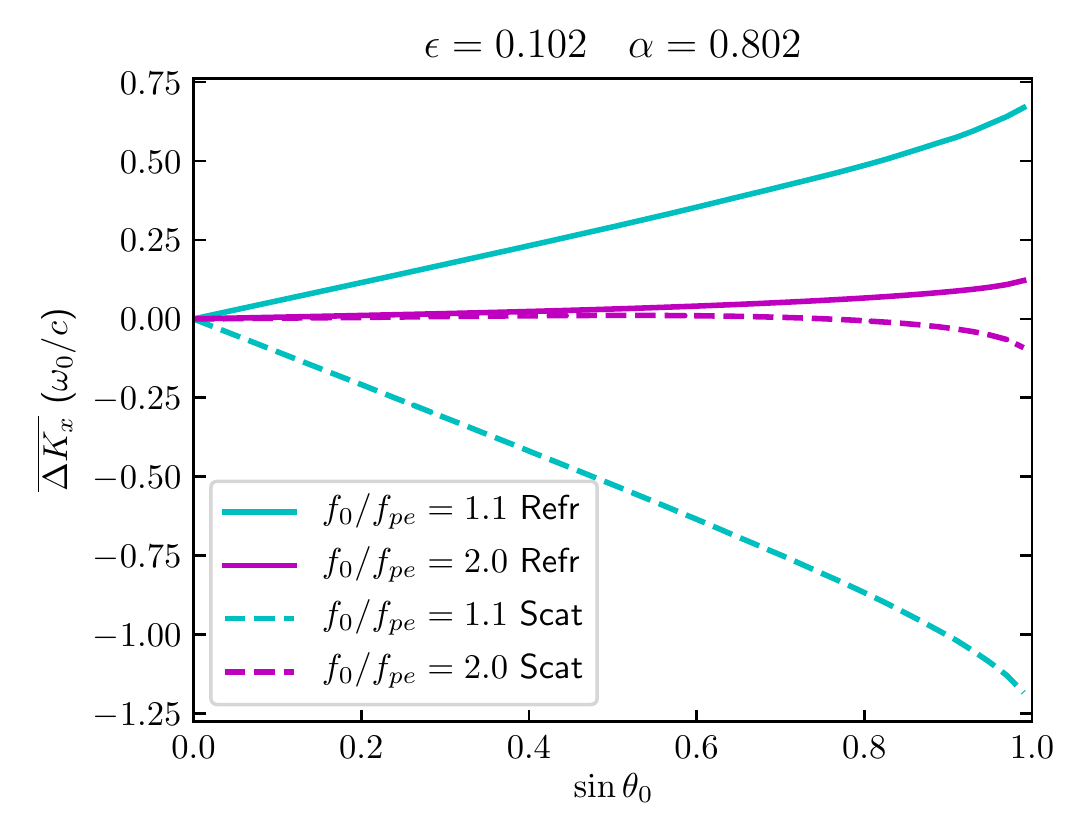}	 
	
	\includegraphics[trim=0.0cm 0cm 0.0cm 0cm, clip=true, width=8.4cm, angle=0]{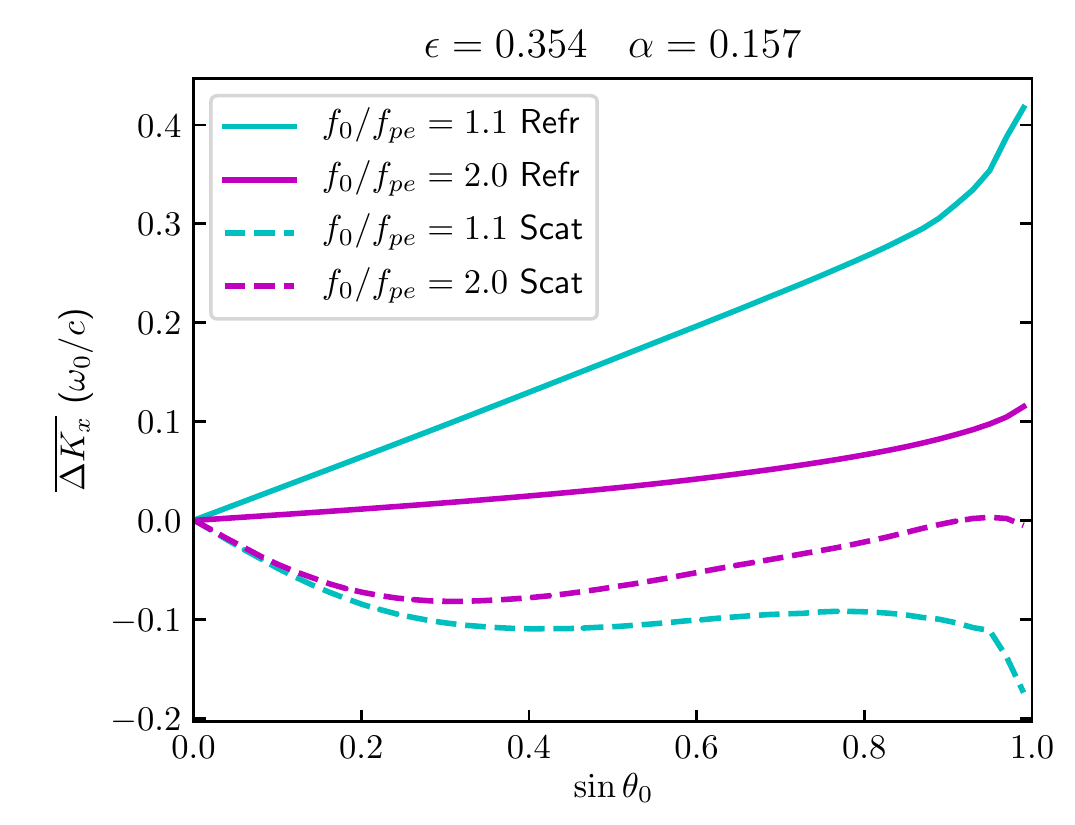}
	\includegraphics[trim=0.0cm 0cm 0.0cm 0cm, clip=true, width=8.4cm, angle=0]{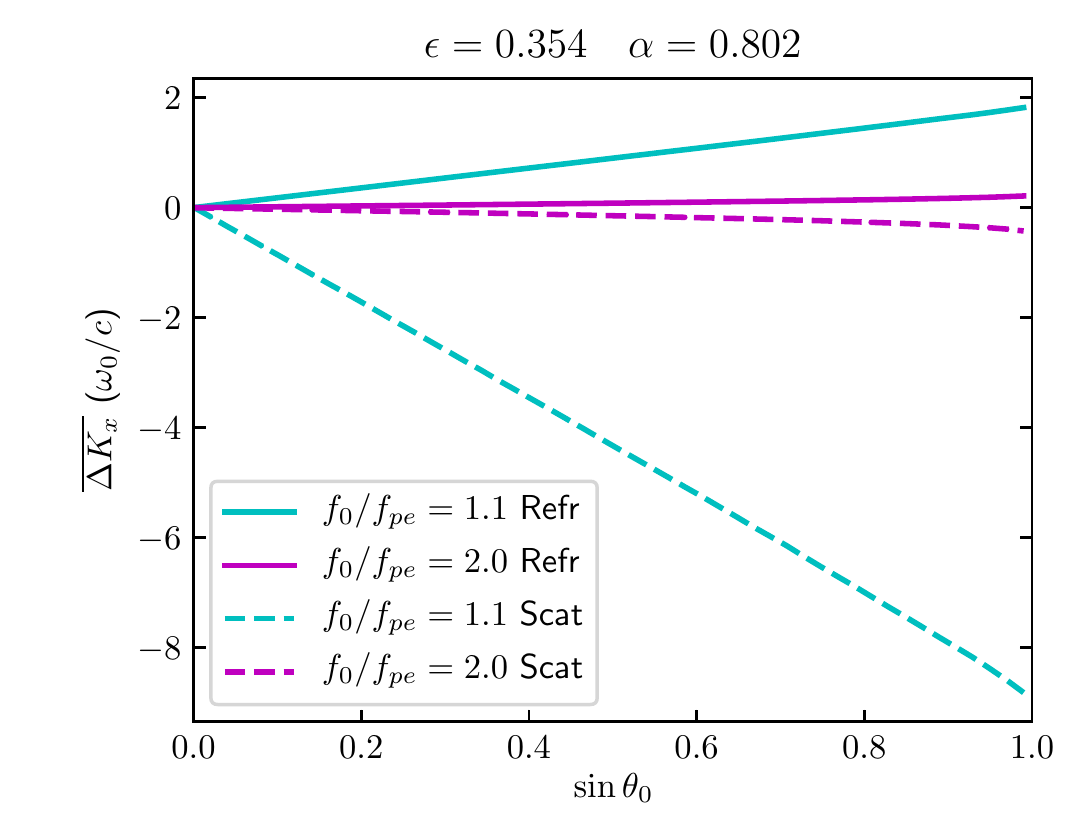} 
	
	\caption{The average value $\overline{\Delta K_x}$ of all collected photons due to influences of scattering (dash lines) and refraction (solid lines) for the waves of fundamental and harmonic emission, respectively. \label{fig:offset_dk}}
\end{figure}

The average change of the wave vector for all photons collected near the observation site can provide a qualitative estimation of the relative influence of scattering and refraction effects on the apparent source position.
In the simulation,
the reconstructed image shows that the source offset is mostly in the $x$ direction in the sky plane. 
Thus, we use $\Delta K_x^l$ to represent the total bending influence of the observed source for the $l\textup{th}$ photon.
For the collected photons in each viewpoint, we calculate the average value $\overline{\Delta K_x}=\Sigma_{l=1}^{l=N}\Delta K_x^l/N$ as a measure  of the influence of the propagation effects on the shift of observed source position. 
Figure \ref{fig:offset_dk} shows the variation of $\overline{\Delta K_x}$ with the position angle of the starting point, where the solid lines and dash lines represent the effects of refraction and scattering,  respectively.

From Figure \ref{fig:offset_dk} one can see that, the contribution of the refraction is in the positive-$x$ direction for the wave vector, while the contribution of scattering is in the negative-$x$ direction. The influences of both refraction and scattering effect are stronger for the fundamental wave than for the harmonic wave in all four cases. When the scattering is more isotropic ($\alpha=0.802$ shown as cyan lines in the right panel), the influence of scattering is overwhelmingly larger than the refraction effect for the fundamental wave, while the influences on the harmonic wave are insignificant.
For high anisotropic scattering cases with small $\alpha$ (left panels in Figure \ref{fig:offset_dk}), the value of $\Delta K_x$ for refraction is comparable to the scattering. 

It needs to note that projecting the wave vector back to the sky plane, a positive change of the wave  vector in the $x$ direction indicates a negative offset of the source position, and \textit{vice versa}. Comparing Figure \ref{fig:offset_dist} and \ref{fig:offset_dk}, one can get that a larger angular scattering rate and  more isotropic scattering would cause the apparent positions of the fundamental and harmonic emission being closer to each other.


\FloatBarrier
\subsection{Source expansion rate and the source speed in the sky plane}
As the evolution of radio observation technology, the radio imaging can be carried out with high time resolution, enabling the study of radio source variation within a solar radio burst, namely the visual speed and expansion rate of the source. 

For the simulation, the \peijin{temporal} variation of the source properties can be obtained from the series of the reconstructed source frame sectioned according to the arrival time of the photons, as shown in Figure \ref{fig:var}. For each parameter set ($\epsilon$, $\alpha$), we linearly fit the time-distance profile of the source centroid for the frames within the flux FWHM in the $x$ direction, the resulting slope of the linear-fit yields the visual speed ($V_x$).
The $V_x$ is calculated in the parameter space of $\epsilon\in[0.03, 0.45]$ and  $\alpha\in[0.05, 0.99]$ with two given starting position angle of $\theta_0=30^\circ$ and $\theta_0=60^\circ$. Accordingly, the angular scattering rate coefficient varies in the space of $\eta\in[8.9\times 10^{-7}, 2.4\times 10^{-4}]\, \textup{km}^{-1}$ for $\epsilon\in[0.03, 0.45]$. 

As shown in Figure \ref{fig:sizet02} and \ref{fig:offset02}, the duration broadening and  position offset due to the wave propagation effects are small for harmonic emission. Therefore, there is large uncertainty in linearly-fitting the source variation trend of harmonic emission. Observation also shows that at a given frequency, the source of the harmonic wave is relatively stable compared with the fundamental wave \citep{zhang2020interferometric}. In the following only variation of the fundamental emission source is presented.

\begin{figure}[hbt!]
	\centering
	\includegraphics[trim=0.0cm 0cm 0.0cm 0cm, clip=true, width=7.5cm, angle=0]{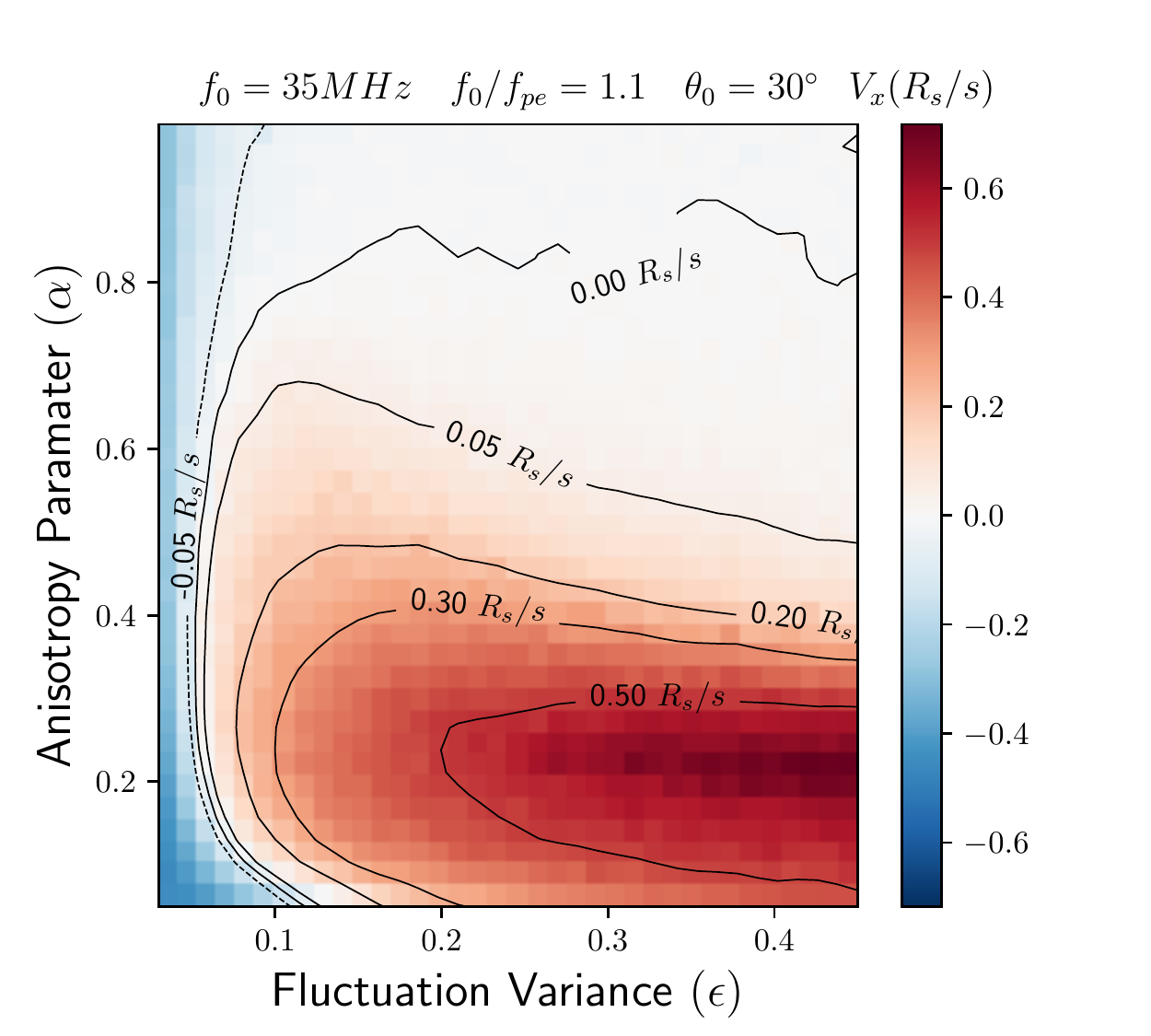}
	\includegraphics[trim=0.0cm 0cm 0.0cm 0cm, clip=true, width=7.5cm, angle=0]{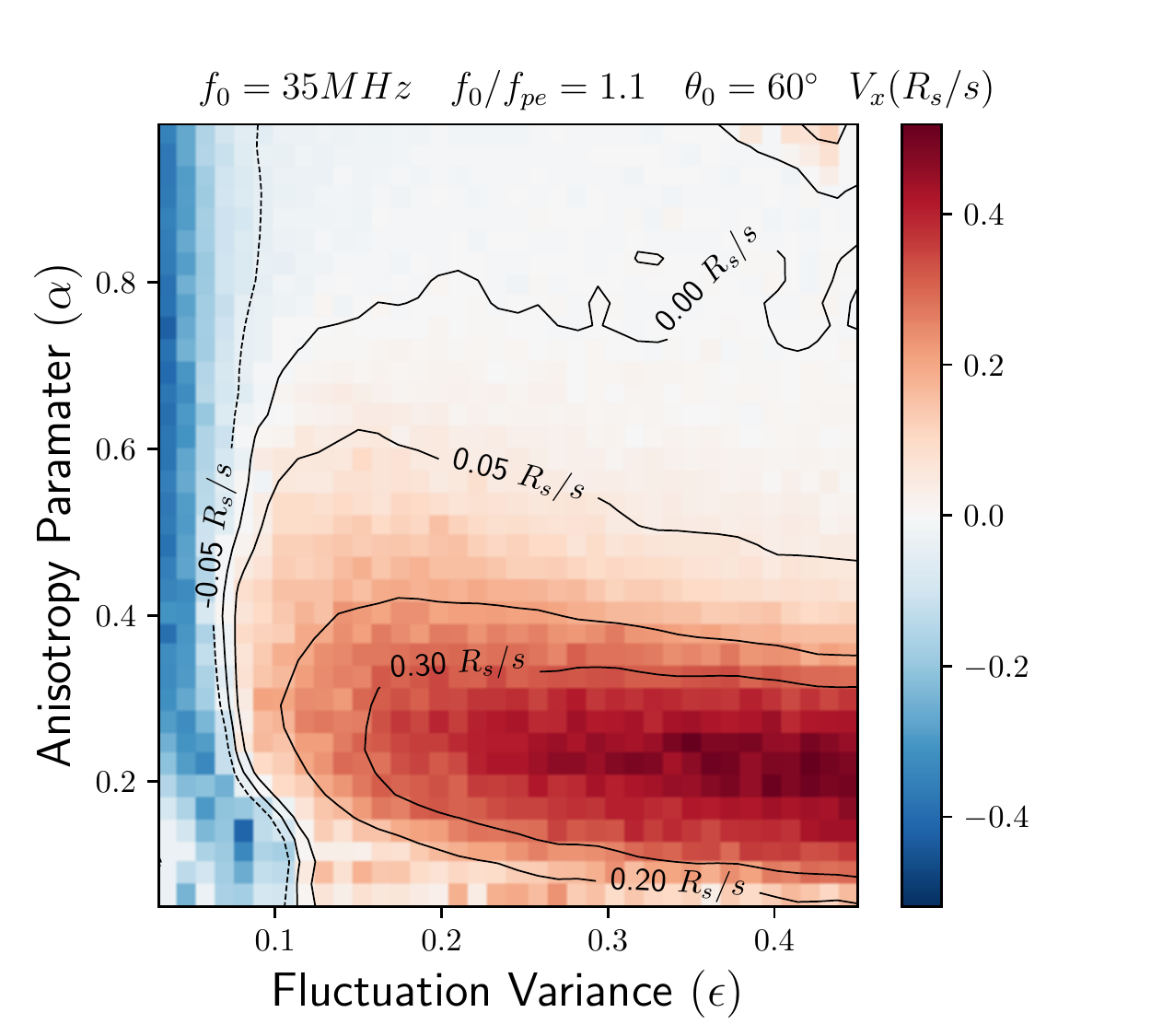}

	\caption{The visual speed of the source for fundamental emission. The left and right panels show the result of the source locations at $\theta_0=30^\circ$ and $\theta_0=60^\circ$. The positive value represents the outward motion from the solar disc center, while the negative value represents the inward motion in this figure. The scale length of density fluctuation $h_0$ is about 860\,km near the wave generation site.}
	\label{fig:vx01}
\end{figure}

Figure \ref{fig:vx01} shows the visual speed of the source of the fundamental emission. The scattering makes the apparent source moving outward from the Sun. 
The patterns of the visual speed are similar for $30^\circ$ and $60^\circ$,
and the major part of the parameter space is divided into two parts. The source tends to have a small visual speed for the more isotropic background ($\alpha>0.5$) and large visual speed requires the background to be highly anisotropic ($\alpha<0.4$). The visual speed reach the maximum at about $\alpha=0.2$ and $\epsilon=0.45$ (corresponds to $\eta=2.4\times10^{-4} \,\textup{km}^{-1}$) within the parameter space for both $30^\circ$ and $60^\circ$ of starting position angle. The maximum visual speed is about $0.65\,R_s/\textup{s}$ or 1.5\,$c$ for $\theta_0 = 30^\circ$ and about $0.5\,R_s/\textup{s}$ or 1.2\,$c$ for $\theta_0 = 60^\circ$. The visual speed of the source with $30^\circ$ starting position angle is larger that of $60^\circ$. Comparing Figure \ref{fig:offset01} and Figure \ref{fig:vx01}, one can find that  direction of offset is mostly identical to the direction of visual speed within the parameter space.

\begin{figure}[hbt!]
    \centering
    \includegraphics[width=8.5cm]{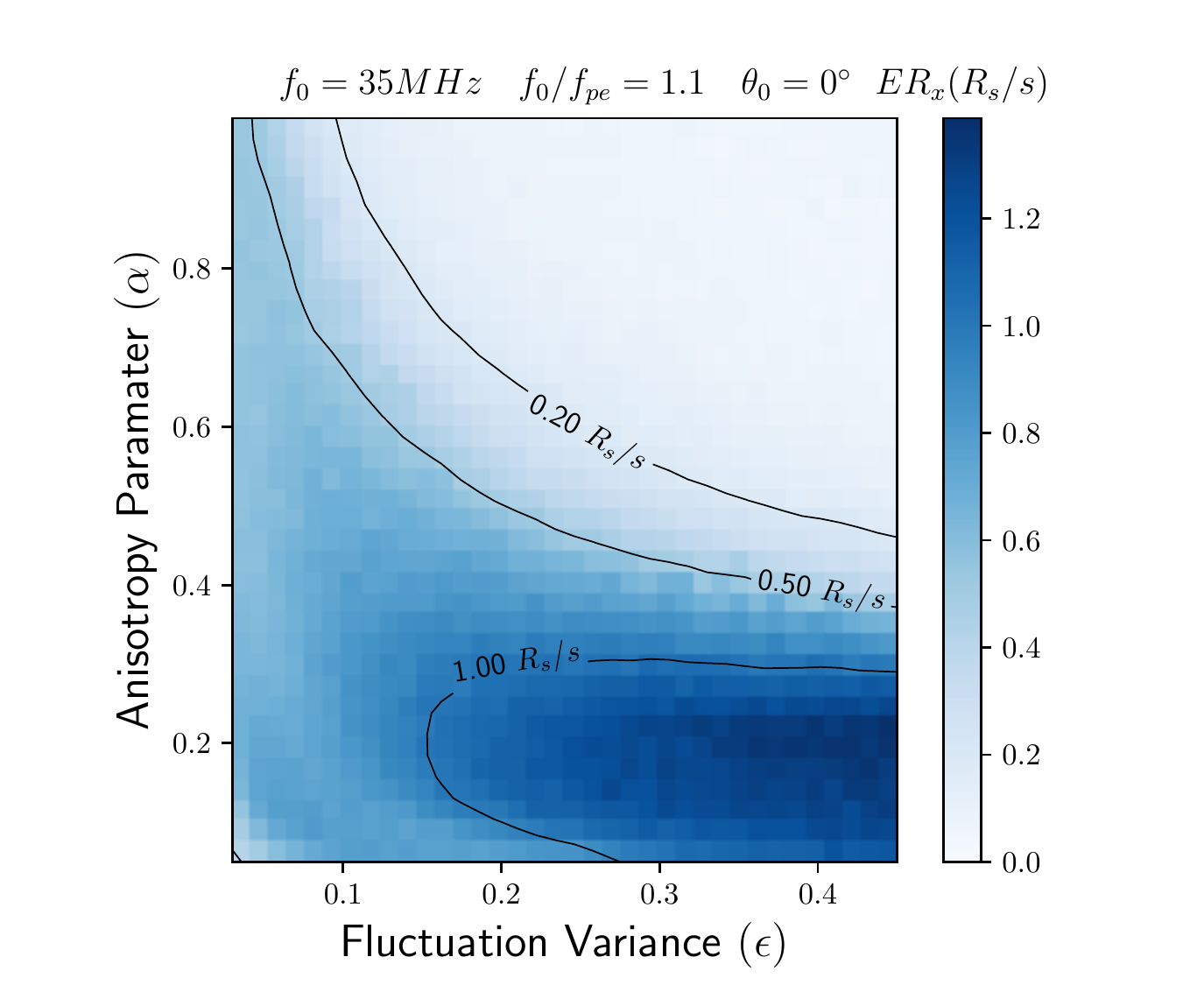}
    \caption{The expansion rate of the source size for fundamental emission located at the solar disk center, measured with the increasing rate of the FWHM width of the visual source.}
    \label{fig:ER}
\end{figure}
For calculating the expansion rate of source size, we use the cases that the original source is located at the solar disk center.
The expansion rate is measured as the slope of linear-fitting the FWHM width in $x$ axis for the flux intensity frames, within the time period  defined by the FWHM range in the time-intensity profile. Figure \ref{fig:ER} shows the expansion rate.
The large expansion rate mainly gathers in the high anisotropic part ($\alpha<0.4$) within the parameter space. The expansion rate can reach $1.4 R_s/\textup{s}$ or $22\,\textup{Arcmin/s}$. Comparing Figure \ref{fig:vx01} and \ref{fig:ER}, one can find that the high expansion rate of the source size is often associated with the large speed of the source apparent motion.  

\FloatBarrier
\section{Discussion}
\peijin{In this study, the source in the calculation is assumed as point pulse source, while the real source could have finite size and duration. Assuming there are no interaction between the electromagnetic waves generated at different time and position, the observed source can be expressed as the convolution of the point pulse response and the time spatial distribution of the real source.}
\peijin{The beam electron generation process at energy release site is considered to be fragmented \citep{benz1994observations,bastian1998radio}, the intrinsic radio sources can have complex spatial and temporal structures. One should be cautious when the simulation results presented in Section 3 are directly compared with the observation results.}
For \peijin{ a burst element of the fragments like a type III burst in} a given frequency, the decay time is determined by the duration of the local excitation decay phase, the velocity dispersion of beam electrons, and the propagation effects \citep{li2008simulations2,ratcliffe2014large,zhang2019source}. The observed size of the source is determined by the real size of the wave excitation regime and the broadening due to the propagation effect \citep{kontar2017imaging}.
Thus, the anisotropic degree and angular scattering rate estimated from the simulation results shown in Figure \ref{fig:sizet01} using Equation (\ref{eq:D}) and (\ref{eq:tau}) are the upper limits. 
The radio burst with fine structures and the short term narrowband radio bursts can provide more constraint on the parameters of the background. For example, the wave generation site of the type IIIb radio burst is believed to be in a compact regime. The apparent source size may be determined mainly by the propagation effects. However, the observed  source size varies largely from case to case,  indicating that the variation of the corona density fluctuations.

The density fluctuation variance and its length scale determine the angular scattering rate of the radio wave. In the simulation of this work, the tuning parameter for the angular scattering rate is $\epsilon$. The length scale is described by the inner scale ($l_i$) and outer scale ($l_o$) of the fluctuation spectrum. Both $l_i$ and $l_o$ are obtained from empirical models \citep{coles1989propagation,wohlmuth2001radio}. 
The inner scale is considered to be the inertial length of ion $d=v_A/\Omega_i$ \citep{coles1989propagation,spangler1990evidence}, where $v_A$ is the local Alfven speed, $\Omega_i$ is the gyro-frequency of ion. The outer scale represents the scale of energy containing in fluctuation. The density fluctuation spectrum and cutoff scales in the corona may be largely different in active region and quiet region. 
Observations from multiple methods like the interplanetary scintillation observations \citep{chang2016ips}, remote and \textit{in situ} observations can help constrain $l_i,l_o$ and $\epsilon$, as well as the angular scattering rate, and benefit the simulation model for the radio wave ray-tracing. Also the radio burst observation combined with simulation results can help to derive the density fluctuation property of the corona \citep{krupar2020density}.

For a given frequency, the wave generation position of fundamental and harmonic emission is at different height according to the plasma emission. While  observational results indicate that the apparent source of fundamental and harmonic emission have co-spatial relationship \peijin{for some type III radio bursts} \citep{dulk1980position}. One possible interpretation on this is that the waves are generated and confined in an density depleted tube \citep{duncan1979wave, Wu2002typeIII}. Alternatively, this may be caused by the refraction and scattering of the waves in their way propagating from the exciting site to the observer \citep{steinberg1971coronal}. In this work, we inspect the source offset and the relative position of the fundamental and harmonic emission of 35\,MHz. The results show that the scattering effect can cause the centroid of the fundamental emission close to that of the harmonic emission. For the four parameter sets shown in Figure \ref{fig:offset_dist}, the visual distance of the reconstructed sources is closer in the case of larger density fluctuation scale. For the case of $\epsilon=0.354$ and $\alpha=0.157$, the position difference $x_H-x_F$ is negative, meaning the source of fundamental emission can appear even slightly higher than the harmonic emission.

The position offset due to the scattering and refraction is not negligible for metric radio bursts generated far away from the solar disk center. The relative error caused by neglecting the wave propagation effects could be up to 50\% for the fundamental and 10\% for the harmonic as shown in Figure \ref{fig:offset_ang_ratio}. The correction of wave propagation effects is essential for the study concerning the position of the source centroid, especially for the fundamental emission. 

From the LOFAR high-cadence imaging spectroscopy, it is found that the apparent sources of a solar type IIIb striae move outward quickly from the Sun with a speed varying from 0.25 to 4 times the speed of light \citep{kontar2017imaging,zhang2020interferometric}. Meanwhile the observed expansion rate of the source size can be as high as $382\pm30 \,\textup{arcmin}^2\textup{s}^{-1}$. A commonly held belief in plasma emission is that type IIIb bursts are generated in a source region with high density inhomogeneity. The simulation results in Figure \ref{fig:vx01} and \ref{fig:ER} show that both the visual speed and the expansion rate can be very large, when the fluctuation variance $\epsilon$ is large and the anisotropy parameter $\alpha$ is small. This is in consistent with the observations and indicates that the density fluctuation may be highly anisotropic in the source region of type IIIb bursts.



\section{Summary}

In this paper, we performed ray-tracing simulation of wave transport for point pulse source. \peijin{For the first time, we explored} the parameter space of the scattering rate coefficient $\eta$ (represented by density fluctuation level $\epsilon$) and the anisotropy parameter $\alpha$ \peijin{with massive number of ray tracing simulation}. 
We analyzed the simulation results to study the influences of wave propagation on the observed source size, position, \peijin{expansion rate, visual speed,} and duration of solar radio bursts \peijin{for both fundamental emission and harmonic emission}. The following are the major conclusions: 
\begin{itemize}
\item
For fundamental emission, both the source size and decay time increase with the scattering rate coefficient or the density fluctuation variance. The isotropic fluctuation can produce much larger decay time than a highly anisotropic fluctuation, while the source size is not sensitive to the level of anisotropy. 
For harmonic emission, both the source size and decay time are largely determined by the density variance. The decay time of harmonic emission is significantly smaller than that of fundamental emission for the same background parameters.

\item 
By comparing the source size and decay time derived from simulation results to the observational statistics, we obtained the estimation of $\eta=8.9\times10^{-5}\, \textup{km}^{-1}$ and $\alpha=0.719$ near the source region of fundamental emission at frequency 35\,MHz.

\item
We derived the position offset of the source and analyzed its dominant factor by decoupling the scattering and refraction in the wave propagation.
The statistical results of source offset show that, the source of the fundamental emission is more outward shifted than that of harmonic emission by the propagation effect, which could account for the co-spatial of fundamental and harmonic emission in \peijin{some} observations.
\item 
The observed source position and size can have significant visual motion and expansion due to the wave propagation effects. 
Both the visual speed and the expansion rate tend to be large for highly anisotropic medium. 
For fundamental emission, the visual speed and the expansion rate for FWHM of the source size can reach respectively about 1.5\,$c$ and 22\,Arcmin/s at $\eta = 2.4\times 10^{-4}\, \mathrm{km^{-1}}$ and $\alpha=0.2$. 
\end{itemize}
A comprehensive comparison of the observed source characteristics and their corresponding values simulated in the parameter space can help us precisely diagnose plasma properties near the wave exciting site of the radio bursts.

In this work, the simulation considers the radio emission of a single frequency (35\,MHz) with the frequency ratio of $f_0/f_{pe}=1.1$ for fundamental emission and $f_0/f_{pe}=2.0$ for harmonic emission. The future work of simulation concerning the parameter space of frequency and frequency-ratio of wave excitation can help to understand the dynamic spectrum of a complete radio burst event and its beam electron property.
The background density model is spherically symmetric in the present work, while the solar corona is a non-uniform medium with a number of large structures. More specific density models with helmet streamers, coronal holes, under-dense or over-dense flux tubes may be also studied in future simulations.

\section{Acknowledgements}
 The research was supported by the National Nature Science Foundation of China (41974199 and 41574167) and the B-type Strategic Priority Program of the Chinese Academy of Sciences (XDB41000000). The numerical calculations in this paper have been done on the supercomputing system in the Supercomputing Center of University of Science and Technology of China.

\bibliography{cite}

\clearpage

\end{document}